\documentclass[a4paper]{iopart}
\usepackage{graphicx}
\usepackage{times}
\usepackage{iopams}

\begin{document}

\title[\texttt{WhiskyMHD}: a new numerical code for general relativistic
  magnetohydrodynamics]{\texttt{WhiskyMHD}: a new numerical code for general
  relativistic magnetohydrodynamics}

\author[B. Giacomazzo \& L. Rezzolla]
{Bruno Giacomazzo$^{1,2}$ and Luciano Rezzolla$^{1,3}$}

\address{$^{1}$ Max-Planck-Institut f\"ur Gravitationsphysik,
  Albert-Einstein-Institut, Golm, Germany}

\address{$^{2}$ SISSA, International School for Advanced Studies and
  INFN, Trieste, Italy}

\address{$^{3}$ Department of Physics, Louisiana State University, Baton Rouge, USA}

\begin{abstract}
  The accurate modelling of astrophysical scenarios involving compact
  objects and magnetic fields, such as the collapse of rotating
  magnetized stars to black holes or the phenomenology of $\gamma$-ray
  bursts, requires the solution of the Einstein equations together
  with those of general-relativistic magnetohydrodynamics. We present
  a new numerical code developed to solve the full set of
  general-relativistic magnetohydrodynamics equations in a dynamical
  and arbitrary spacetime with high-resolution shock-capturing
  techniques on domains with adaptive mesh refinements. After a
  discussion of the equations solved and of the techniques employed,
  we present a series of testbeds carried out to validate the code and
  assess its accuracy. Such tests range from the solution of
  relativistic Riemann problems in flat spacetime, over to the
  stationary accretion onto a Schwarzschild black hole and up to the
  evolution of oscillating magnetized stars in equilibrium and
  constructed as consistent solutions of the coupled Einstein-Maxwell
  equations.
\end{abstract}


\section{\label{intro}Introduction}

Magnetic fields are ubiquitous in astrophysical objects and can play
an important role, especially in those scenarios involving compact
objects such as neutron stars and black holes. An accurate and
consistent modelling of these scenarios, which are extreme both for
the gravitational and the electromagnetic fields, cannot be done
analytically and perturbative methods are also of limited validity. In
the absence of symmetries, in fact, no dynamical and analytic
solutions are known and it is only through the full solution of the
equations of general-relativistic magnetohydrodynamics (GRMHD) that
one can hope to improve our knowledge of these objects under realistic
conditions.

As in general-relativistic hydrodynamics, the work in this area of
research has started, more than 30 years ago with the pioneering work
of Wilson~\cite{wilson75} in lower spatial dimensions
(see~\cite{font03} for a review of the technical and scientific
progress in relativistic hydrodynamics). Unlike general relativistic
hydrodynamics, however, where both technical issues and scientific
investigations have now reached an advanced stage of sophistication
and accuracy, progress in GRMHD has yet to reach a comparable level of
maturity.  Indeed, it was only over the last few years that the slow
but steady progress in GRMHD has seen a renewed burst of activity,
with a number of groups developing a variety of numerical codes
solving the equations of GRMHD under different approaches and
approximations. This is partly due to the considerable added
complexity of the set of equations to be solved in GRMHD and, partly,
to the fact that only recently sufficient computational resources have
become available to tackle this problem in two or three spatial
dimensions and with sufficient resolution.

It does not come as a surprise, therefore, that most of the numerical
codes developed in the last decade have been based on the same
non-conservative formulation of the GRMHD equations introduced by
Wilson, solving them on a fixed background to study accretion disks
around black holes. In~\cite{yokosawa93,yokosawa95}, for instance, the
effects of a Kerr black hole on magnetohydrodynamical accretion have
been studied, with particular attention being paid to the transfer of
energy and angular momentum. Koide {\it et al.}~\cite{koide99}, on the
other hand, have developed a numerical code based on the
artificial-viscosity approach proposed by Davis~\cite{davis87} to
perform the first simulations of jet formation in General
Relativity~\cite{koide00} and to study the possibility of extracting
the rotational energy from a Kerr black
hole~\cite{koide02,koide03}. Furthermore, a distinct numerical code
has been constructed by De Villiers and Hawley~\cite{devilliers03}
using the formulation proposed in~\cite{hws84a,hws84b} to carry out a
series of studies on accretion flows around Kerr black
holes~\cite{devilliers03b,hirose04,devilliers05}.

It is only rather recently that different groups have started to
recast the system of GRMHD equations into a conservative form in order
to benefit of the use of high-resolution shock-capturing schemes
(HRSC)~\cite{harm,komissarov04,anton06,neilsen06,monteroetal07}. Such
schemes, we recall, are essential for a correct representation of
shocks, whose presence is expected in several astrophysical scenarios
and in particular in those involving compact objects. Two mathematical
results corroborate this view, with the first one stating that a
stable scheme converges to a weak solution of the hydrodynamical
equations~\cite{Lax60}, and with the second one showing that a
non-conservative scheme will converge to the wrong weak solution in
the presence of a shock~\cite{Hou94}.

All of these newly developed codes~\cite{harm, komissarov04, anton06,
  neilsen06} that make use of HRSC methods have been so far applied to
the study of accretion problems onto black holes, for which the
self-gravity of the accreting material introduces very small
corrections to the spacetime and fixed spacetime backgrounds can be
used satisfactorily.

Approaches alternative to that of constructing GRMHD codes have
instead been based on the use of a modified Newtonian gravitational
potential to mimic general relativistic effects without having to
solve numerically Einstein equations (see~\cite{obergaulinger06a} for
an application to magnetorotational collapse of stellar cores) or on
the use of different numerical methods, such as smoothed particle
hydrodynamics and artificial viscosity, to study the merger of binary
neutron star systems as a possible engine for short $\gamma$-ray
bursts~\cite{price06}. Although the use of these approximations has
made it possible to investigate this astrophysical scenario for the
first time including details about the microphysics, it is clear that
equally important corrections coming from the dynamical evolution of
the spacetime need to be introduced when trying to model the
phenomenology that is thought to be behind $\gamma$-ray bursts.  As a
first step in this direction, two codes were recently developed to
solve the full set of GRMHD equations on a dynamical
background~\cite{duez05,shibata05}. These codes, in particular, were
used to perform the first study, in two spatial dimensions, of the
collapse of magnetized differentially rotating neutron
stars~\cite{duez06a,duez06b,duez06c} which are thought to be good
candidates for short $\gamma$-ray bursts.

Here, we present \texttt{WhiskyMHD}, a new three-dimensional numerical
code in Cartesian coordinates developed to solve the full set of GRMHD
equation on a dynamical background. The code is based on the use of
high-resolution shock-capturing techniques on domains with adaptive
mesh refinements, following an approach already implemented with
success in the general-relativistic hydrodynamics code
\texttt{Whisky}~\cite{Baiotti03a}, and which has been used in the
study of several astrophysical scenario with particular attention to
gravitational-wave emission from compact objects.

The paper is organized as follows: in Sect.~\ref{equations} we recall
the equations of GRMHD and the form they assume when recast in a
conservative form, while in Sect.~\ref{sec:numerical} we discuss in
detail the numerical methods adopted for their
solution. Section~\ref{sec:tests} is dedicated to the series of
testbeds the code has passed both in special and in general
relativistic conditions. Finally, Sect.~\ref{conclusion} offers a
summary of the results and an overview on our future projects.

Throughout the paper we use a spacelike signature $(-,+,+,+)$ and a
system of units in which $c=G=M_\odot =1$. Greek indices are taken to
run from 0 to 3, Latin indices from 1 to 3 and we adopt the standard
convention for the summation over repeated indices. Finally we
indicate 3-vectors with an arrow and use bold letters to denote
4-vectors and tensors.

\section{\label{equations}Formulation of the equations}

We adopt the usual 3-dimensional foliation of the spacetime so that
the line element reads
\begin{equation}
\mathrm{d} s^2 = -(\alpha^2 - \beta^i\beta_i) \mathrm{d} t^2 + 2\beta_i\mathrm{d} x^i\mathrm{d} t
+ \gamma_{ij}\mathrm{d} x^i\mathrm{d} x^j\ ,
\end{equation}
where $\beta^i$ is the {\it shift} vector, $\alpha$ is the {\it lapse}
function and $\gamma_{ij}$ are the spatial components of the
four-metric $g_{\mu\nu}$.

As its predecessor \texttt{Whisky}~\cite{Baiotti03a}, the
\texttt{WhiskyMHD} code benefits of the {\tt Cactus} computational
toolkit~\cite{Cactusweb} which provides an infrastructure for the
parallelization and the I/O of the code, together with several methods
for the solution of the Einstein equations. As a result, at each
timestep our new code solves the MHD equations while Cactus provides
the evolution of the metric quantities. The evolution of the field
components is done using the NOK
formulation~\cite{Nakamura87,shibata95,baumgarte99} and details about
its numerical implementation can be found
in~\cite{Alcubierre99d,Alcubierre02a, whisky}.

Here too we make use of the so-called ``Valencia formulation''
\cite{Marti91,Banyuls97} which was originally developed as a $3+1$
conservative Eulerian formulation of the general relativistic
hydrodynamic equations, but which has been recently extended to the case
of GRMHD~\cite{anton06}. Following~\cite{anton06} we define the Eulerian
observer as the one moving with four velocity $\mathbf{n}$ perpendicular
to the hypersurfaces of constant $t$ at each event in the spacetime. This
observer measures the following three-velocity of the fluid
\begin{equation}
v^i = \frac{h^i_\mu u^\mu}{-u^\mu n_\mu} = \frac{u^i}{W}+\frac{\beta^i}{\alpha} \;,
\end{equation}
where $h_{\mu \nu} \equiv g_{\mu \nu}+n_\mu n_\nu $ is the projector
orthogonal to ${\boldsymbol n}$, ${\boldsymbol u}$ is the
four-velocity of the fluid and $-u^\mu n_\mu=\alpha u^0=W$ is the
Lorentz factor which satisfies the usual relation $W=1/\sqrt{1-v^2}$,
where $v^2\equiv \gamma_{ij}v^iv^j$. The covariant components of the
three-velocity are simply given by $v_i=u_i/W$.

\subsection{\label{sec:maxwell}Maxwell equations}
The electromagnetic field is completely described by the Faraday
electromagnetic tensor field $F^{\mu \nu}$ obeying Maxwell equations
(cfr~\cite{anile})
\begin{eqnarray}
\nabla_\nu\, ^{*\!\!}F^{\mu \nu} &=& 0 \;,\\
\nabla_\nu F^{\mu \nu}     &=& 4 \pi \mathcal{J}^\mu \;,
\end{eqnarray}
where $\nabla$ is the covariant derivative with respect to the
four-metric ${\boldsymbol g}$, ${\boldsymbol {\mathcal J}}$ is the charge
current four-vector and $\,^{*\!\!}{\boldsymbol F}$ is the dual of the
electromagnetic tensor defined as
\begin{equation}
\,^{*\!\!}F^{\mu \nu} = \frac{1}{2} \eta^{\mu \nu \lambda \delta} F_{\lambda \delta} \;,
\end{equation}
$\eta^{\mu \nu \lambda \delta}$ being the Levi-Civita pseudo-tensor. A
generic observer with four-velocity ${\boldsymbol U}$ will measure a
magnetic field ${\boldsymbol B}$ and an electric field ${\boldsymbol E}$
given by
\begin{eqnarray}
E^\alpha &\equiv& F^{\alpha \beta} U_\beta \;, \\
B^\alpha &\equiv& \,^{*\!\!}F^{\alpha \beta} U_\beta \;,
\end{eqnarray}
and the charge current four-vector ${\boldsymbol {\mathcal J}}$ can be in
general expressed as
\begin{equation}
\mathcal{J}^\mu = q u^\mu + \sigma F^{\mu \nu} u_\nu \;,
\end{equation}
where $q$ is the proper charge density and $\sigma$ is the electric
conductivity. 

Hereafter we will assume that our fluid is a perfect conductor (ideal MHD
limit) and so that $\sigma \rightarrow \infty$ and $F^{\mu \nu} u_\nu=0$
(i.e. the electric field measured by the comoving observer is zero) in
order to keep the current finite. In this limit, the electromagnetic
tensor and its dual can be written exclusively in terms of the magnetic
field ${\boldsymbol b}$ measured in the comoving frame
\begin{equation}
F^{\nu \sigma} = \eta^{\alpha\mu\nu\sigma}b_\alpha u_\mu \;,
\hskip 1.0cm
\,^{*\!\!}F^{\mu\nu} = b^\mu u^\nu - b^\nu u^\mu \; .
\end{equation}
with the Maxwell equations taking the simple form
\begin{equation}
\nabla_\nu \,^{*\!\!}F^{\mu\nu} = \frac{1}{\sqrt{-g}} 
\partial_\nu \left( \sqrt{-g}
\left(b^\mu u^\nu - b^\nu u^\mu \right)\right) = 0 \; ,
\label{eq:maxwell}
\end{equation}
In order to express these equations in terms of quantities measured by an
Eulerian observer, we need to compute the relation between the magnetic
field measured by the comoving and by the Eulerian observers,
respectively ${\boldsymbol b}$ and ${\boldsymbol B}$. To do that we
introduce the projection operator $P_{\mu\nu}\equiv g_{\mu\nu}+u_\mu
u_\nu$ orthogonal to ${\boldsymbol u}$. If we apply this operator to the
definition of the magnetic field ${\boldsymbol B}$ measured by an
Eulerian observer, we can easily derive the following relations
\begin{equation}
b^0 = \frac{W B^i v_i}{\alpha} \;,  
b^i = \frac{B^i + \alpha b^0 u^i}{W} \;,
b^2 \equiv b^\mu b_\mu = \frac{B^2 + \alpha^2 (b^0)^2}{W^2} \;, 
\label{eq:b0b2}
\end{equation}
where $B^2\equiv B^i B_i$.
The time component of equations \eref{eq:maxwell} provides the
divergence-free condition
\begin{equation}
\partial_i \tilde{B}^i =0 \;,
\label{eq:divergence}
\end{equation}
where $\tilde{B}^i\equiv \sqrt{\gamma}B^i$ and $\gamma$ is the
determinant of $\gamma_{ij}$. The spatial components of equations
\eref{eq:maxwell}, on the other hand, yield the induction equations
for the evolution of the magnetic field
\begin{equation}
\partial_t (\tilde{B}^i) = 
\partial_j(\tilde{v}^i\tilde{B}^j-\tilde{v}^j\tilde{B}^i) \;,
\label{eq:induction}
\end{equation}
where $\tilde{v}^i\equiv \alpha v^i-\beta^i$.

\subsection{\label{sec:fluid}Conservation equations}

The evolution equations for the rest-mass density $\rho$, the specific
internal energy $\epsilon$ and for the three-velocity $v^i$ can be
computed, as done in relativistic hydrodynamics, from the conservation of
the baryon number
\begin{equation}
\nabla_\nu (\rho u^\nu) =0 \;,
\label{eq:baryon}
\end{equation}
from the conservation of the energy-momentum
\begin{equation}
\nabla_\nu T^{\mu\nu} = 0 \;,
\label{eq:energy-momentum}
\end{equation}
and from an equation of state (EOS) relating the gas pressure $p$ to the
rest-mass density $\rho$ and to the specific internal energy $\epsilon$.
We also assume that the fluid is perfect so that the total energy
momentum tensor, including the contribution from the magnetic field, is
given by
\begin{equation}
T^{\mu\nu} = \left( \rho h + b^2\right) u^\mu u^\nu +
\left(p+\frac{b^2}{2}\right)g^{\mu\nu}-b^\mu b^\nu \; ,
\end{equation}
where $h=1+\epsilon+p/\rho$ is the specific relativistic enthalpy.

Following~\cite{anton06} and in order to make use of HRSC methods, we
rewrite equations \eref{eq:baryon}, \eref{eq:energy-momentum} and
\eref{eq:induction} in the following conservative form
\begin{equation}
\frac{1}{\sqrt{-g}} \left[\partial_t(\sqrt{\gamma}\mathbf{F}^0) +
\partial_i(\sqrt{-g}\mathbf{F}^i)\right]=\mathbf{S} \;,
\label{eq:conservative}
\end{equation}
where $\mathbf{F}^0 \equiv (D, S_j, \tau, B^k)^{\mathrm{T}}$ is the
vector of the conserved variables measured by the Eulerian observer,
$\mathbf{F}^i$ are the fluxes
\begin{equation}
  \mathbf{F}^i = \left(
  \begin{array}{c}
    D{\tilde{v}^i}/{\alpha} \\\nonumber\\
    S_j{\tilde{v}^i}/{\alpha}+\left(p+{b^2}/{2}\right)
    \delta^i_j-b_j{B^i}/{W}\\\nonumber\\
    \tau{\tilde{v}^i}/{\alpha}+\left(p+{b^2}/{2}\right)v^i-
    \alpha b^0 {B^i}/{W}\\\nonumber\\
    {B^k \tilde{v}^i}/{\alpha}  - B^i {\tilde{v}^k}/{\alpha}  
  \end{array}
  \right) \; ,
\end{equation}
and $\mathbf{S}$ are the source terms
\begin{equation}
  \mathbf{S} = \left(
  \begin{array}{c}
    0\\\nonumber\\
    T^{\mu\nu}\left(\partial_\mu g_{\nu j}-
    \Gamma^\delta_{\nu\mu}g_{\delta j}\right)\\\nonumber\\
    \alpha\left(T^{\mu 0}\partial_\mu \ln\alpha - 
    T^{\mu\nu}\Gamma^0_{\nu\mu}\right)\\\nonumber\\
    0^k
  \end{array}
  \right) \; ,
\end{equation}
where 
\begin{eqnarray}
D    & \equiv & \rho W \; , \label{defD} \\
S_j  & \equiv & (\rho h + b^2)W^2 v_j -\alpha b^0 b_j \; , \label{defSj} \\
\tau & \equiv & (\rho h + b^2)W^2 - (p+\frac{b^2}{2})-\alpha^2(b^0)^2-D \; , \label{deftau} 
\end{eqnarray}
and $0^k=(0,0,0)^{\mathrm{T}}$. While ready to make use of arbitrary EOS,
the ones presently implemented in the code have a rather simple form and
are given by either the polytropic EOS
\begin{eqnarray}
\label{poly}
p &=& K \rho^{\Gamma}\ , \\
\epsilon &=& \frac{p}{\rho (\Gamma-1)}\ ,
\end{eqnarray}
or by the ideal-fluid EOS
\begin{equation}
\label{id_fluid}
p = (\Gamma-1) \rho\, \epsilon \ ,
\end{equation}
where $K$ is the polytropic constant and $\Gamma$ is the adiabatic
exponent. In the case of the polytropic EOS~\eref{poly}, $\Gamma=1+1/N$,
where $N$ is the polytropic index.

\section{\label{sec:numerical}Numerical methods}

As in the \texttt{Whisky} code, the evolution equations are integrated in
time using the method of lines~\cite{toro99}, which reduces the partial
differential equations~\eref{eq:conservative} to a set of ordinary
differential equations that can be evolved using standard numerical
methods, such as Runge-Kutta or the iterative Cranck-Nicholson
schemes~\cite{teukolsky,leiler}.

\subsection{\label{sec:hlle}Approximate Riemann solver}

As mentioned in the Introduction, \texttt{WhiskyMHD} makes use of HRSC
schemes based on the use of Riemann solvers to compute the fluxes between
the numerical cells. More specifically, we have implemented the
Harten-Lax-van~Leer-Einfeldt (HLLE) \textit{approximate} Riemann
solver~\cite{hlle} which is simply based on the calculation of the
eigenvalues of eqs.~\eref{eq:conservative} and it does not require a
basis of eigenvectors.  In the HLLE formulation the flux at the interface
between two numerical cells is therefore computed as
\begin{equation}
  \mathbf{F}^i=\frac{c_{min}\mathbf{F}^i_r+c_{max}\mathbf{F}^i_l-c_{max}c_{min}
    \left(\mathbf{F}^0_r-\mathbf{F}^0_l\right)}{c_{max}+c_{min}} \; ,
  \label{eq:hlleflux}
\end{equation}
where $\mathbf{F}^\mu_r$ and $\mathbf{F}^\mu_l$ are computed from the
values of the primitive variables reconstructed at the right and left
side of the interface ${\boldsymbol P}_r$ and ${\boldsymbol P}_l$,
respectively. The coefficients $c_{max}\equiv \max(0,c_{+,r},c_{+,l})$,
$c_{min}\equiv -\min(0,c_{-,r},c_{-,l})$ and $c_{\pm,r}$, $c_{\pm,l}$ are
instead the maximum left- and right-going wave speeds computed from
${\boldsymbol P}_r$ and ${\boldsymbol P}_l$. In our implementation
${\boldsymbol P}_r$ and ${\boldsymbol P}_l$ are computed using a second
order TVD slope limited method which can be used with different limiters
such as minmod, Van Leer and MC~\cite{toro99}.

An alternative to the use of an \textit{approximate} Riemann solver
could have been the use of the \textit{exact} Riemann solver recently
developed in GRMHD. We recall that in relativistic hydrodynamics the
exact solution is found after expressing all of the quantities behind
each wave as functions of the value of the unknown gas pressure $p$ at
the contact discontinuity. In this way, the problem is reduced to the
search for the value of the pressure that satisfies the jump
conditions at the contact discontinuity
(see~\cite{marti94,pons00,rezzolla01,rezzolla03} for the details). The
procedure for the exact solution of the Riemann problem in
relativistic MHD is based instead on the use of an hybrid approach
that makes use of different set of unknowns depending on the wave. In
the case of fast-magnetosonic waves, both shocks or rarefactions, all
the variables behind the waves are rewritten as functions of the total
pressure, {\it i.e.}  $p+b^2/2$, while behind slow magnetosonic shocks
or rarefactions the components of the magnetic field tangential to the
discontinuity are used to compute all the other variables. The use of
this strategy was essential in order to reduce the problem to the
solution of a closed system of equations that can be solved with
standard numerical routines such as Newton-Raphson schemes
(see~\cite{giacomazzo06} for the details). The numerical code
computing the exact solution is freely available from the authors upon
request and it is now becoming a standard tool in the testing of both
special and general relativistic MHD codes.

While the use of an exact Riemann solver could provide the solution of
the discontinuous flow at each cell interface with arbitrary accuracy,
the exact solver presented in~\cite{giacomazzo06} is still
computationally too expensive to be used in ordinary multidimensional
codes and we have found the HLLE algorithm to be sufficiently accurate
for the resolution used in our tests.

\subsubsection{Calculation of the eigenvalues}
\label{whiskymhd_eigenvalues}

~~~An important difference with respect to relativistic-hydrodynamic
codes is that in GRMHD the calculation of the eigenvalues required by the
HLLE solver is made more complicated by the solution of a quartic
equation. The characteristic structure of GRMHD equations is analyzed in
detail in~\cite{anile} and we simply report here the expressions for the
calculation of the seven wave speeds associated with the entropic, the
Alfv\'en, the fast and slow magnetosonic waves.

More specifically, the characteristic speed $\lambda$ of the entropic
waves is simply given by $\tilde{v} = \alpha v^i-\beta^i$, while the
values for the left- and right-going Alfv\'en waves are
\begin{equation}
\lambda^i_{1,2} = \frac{b^i \pm \sqrt{(\rho h + b^2)} u^i}{b^0 \pm
\sqrt{(\rho h + b^2)} u^0} \; .
\end{equation}

Similarly, the four speeds that are associated with the fast and slow
magnetosonic waves, and that are required in the calculation of the
fluxes, can be obtained by the solution of the following quartic equation
in each direction $i$ for the unknown $\lambda$
\begin{equation}
\rho h \left(\frac{1}{c_s^2}-1\right)a^4 -
\left(\rho h + \frac{b^2}{c_s^2}\right)a^2 G + {\cal B}^2 G = 0 \; ,
\label{eq:quartic}
\end{equation}
where
\begin{eqnarray}
a &\equiv& \frac{W}{\alpha} \left(-\lambda+ \alpha v^i - \beta^i \right) \; ,\\
\cal{B} &\equiv& b^i - b^0 \lambda \; , \\
G &\equiv& \frac{1}{\alpha^2}\left[-(\lambda+\beta^i)^2+\alpha^2\gamma^{ii}\right] \; ,
\end{eqnarray}
and $c_s$ is the sound speed (Note that the convention on repeated
indices should not be used for the last term in the expression for $G$,
i.e. $\gamma^{ii}$). In the degenerate case in which $B^i=0$,
eq.~\eref{eq:quartic} can be reduced to a simple quadratic equation that
is solved analytically. In the more general case, however,
eq.~\eref{eq:quartic} cannot be reduced to the product of two quadratic
equations as in Newtonian MHD and different methods are implemented in
the code in order to solve it. The first one simply makes use of the
analytic expression for a quartic equation~\cite{abram}, while the other
two search the solution numerically either through an eigenvalue method
or through a Newton-Raphson iteration~\cite{NR}. The latter has shown to
be the most accurate and robust and it is the one used by default.

We have also implemented an approximate method for the calculation of the
eigenvalues associated with the fast magnetosonic waves (which are the
only two roots needed by HLLE) which was introduced in~\cite{leismann}
and which reduces the original quartic to a quadratic equation, that can
be solved analytically, by imposing $B^i=0$ and $B^j v_j=0$ in
equation~\eref{eq:quartic}. The values computed in this way differ by
less than $1\%$ with respect to the exact values and we have used them in
those situations in which the solution of the quartic can be complicated
by the presence of degeneracies or when two of the roots are very close
to each other.

\subsection{\label{sec:CT}Constrained-Transport Scheme}

Although an exact solution of eqs.~\eref{eq:induction} would guarantee
that the constraint condition~\eref{eq:divergence} is also satisfied
identically and all times, any numerical solution of the induction
equations~\eref{eq:induction} will inevitably produce a violation of
the divergence-free condition which, in turn, may lead to unphysical
results or even to the development of
instabilities~\cite{brackbill80}. To avoid this problem several
numerical methods were developed in the past starting from the
so-called ``staggered mesh magnetic field transport algorithm'' first
proposed by Yee~\cite{Yee} and then implemented in an
artificial-viscosity scheme with the name of ``constrained-transport''
scheme (CT) by {Evans} \& {Hawley}~\cite{evans88}.

A modified version of the CT scheme~\cite{evans88} which is based on the
use of the fluxes computed with HRSC methods has been proposed by Balsara \&
Spicer~\cite{balsara99} (``flux-CT'') and is the one implemented in our
code because of its simplicity and computational efficiency.  An
interested reader is referred to~\cite{toth00} for other possible methods
to enforce the divergence-free condition with HRSC schemes.

\begin{figure}
  \begin{center}
    \includegraphics[width=0.6\textwidth]{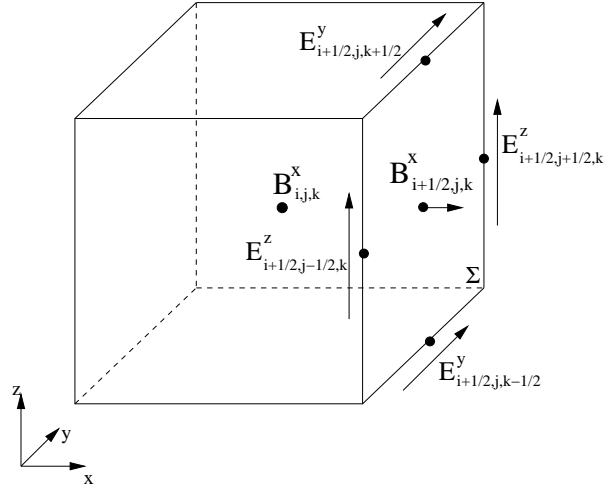}
  \end{center}
  \caption{\label{fig:divB}Schematic diagram of the data needed for
    the CT scheme. The evolution of $\tilde{B}^x_{i+1/2,j,k}$ is
    determined by the values of the electric field $\vec{E}$ at the
    edges of the surface $\Sigma$ located at $(i+1/2,j,k)$.}
\end{figure}

We recall that the flux-CT method is based on the relation that exists in
ideal MHD between the fluxes of the magnetic field $\vec{B}$ and the
value of the electric field $\vec{E}\equiv
-\vec{\tilde{v}}\times\vec{\tilde{B}}$. In particular, if we define
$\mathbf{\tilde{F}}^i\equiv \alpha\sqrt{\gamma}\mathbf{F}^i$ then the
following relations hold
\begin{eqnarray}
  E^x &= \tilde{F}^z\left(\tilde{B}^y\right) &= -\tilde{F}^y\left(\tilde{B}^z\right) \label{eq:Exflux} \; ,\\
  E^y &= -\tilde{F}^z\left(\tilde{B}^x\right)  &= \tilde{F}^x\left(\tilde{B}^z\right) \label{eq:Eyflux} \; ,\\
  E^z &= \tilde{F}^y\left(\tilde{B}^x\right) &= -\tilde{F}^x\left(\tilde{B}^y\right) \label{eq:Ezflux} \; ,
\end{eqnarray}
where $\tilde{F}^i\left(\tilde{B}^j\right)\equiv \tilde{v}^i\tilde{B}^j -
\tilde{v}^j\tilde{B}^i$.
The induction equation~\eref{eq:induction} can then be written as
\begin{equation}
\partial_t \vec{\tilde{B}} + \vec{\nabla}\times\vec{E} = 0 \; .
\label{eq:CTinduction}
\end{equation}
Taking the surface integral of~(\ref{eq:CTinduction}) across a surface
$\Sigma$ between two numerical cells, Stokes' theorem yields
\begin{equation}
  \partial_t \int_\Sigma \vec{\tilde{B}}\cdot\mathrm{d}\vec{\Sigma}
  +\int_{\partial\Sigma} \vec{E} \cdot \vec{l}=0 \; ,
  \label{eq:CTintegral} 
\end{equation}
where $\vec{l}$ is the unit vector parallel to the surface boundary
$\partial\Sigma$. Considering for simplicity the $x$-direction, with
the surface $\Sigma$ as located at $(i+\frac{1}{2},j,k)$  and the
integers $(i,j,k)$ denoting the cell centers on our discrete grid
(see~\fref{fig:divB}), we can define
\begin{equation}
  \tilde{B}^x_{i+\frac{1}{2},j,k}\equiv \frac{1}{\Delta y \Delta z}\int_\Sigma
  \vec{\tilde{B}}\cdot\mathrm{d}\vec{\Sigma} \; ,
\end{equation}
and use the finite-difference expression of eq.~\eref{eq:CTintegral} to
obtain
\begin{eqnarray}
&&\fl  \partial_t \tilde{B}^x_{i+\frac{1}{2},j,k} = -  
\left(E^y_{i+\frac{1}{2},j,k-\frac{1}{2}}-E^y_{i+\frac{1}{2},j,k+
\frac{1}{2}}\right)/{\Delta z} -  
\left(E^z_{i+\frac{1}{2},j+\frac{1}{2},k}-E^z_{i+\frac{1}{2},j-
\frac{1}{2},k}\right)/{\Delta y} \; , \nonumber\\
 \label{eq:CTdiscrete}
\end{eqnarray}
where the values of the electric field on the edges of the surface are
simply computed taking the arithmetic mean of the fluxes across the
surfaces that have that edge in common
[cf~\eref{eq:Exflux}--\eref{eq:Ezflux}], e.g.
\begin{equation}
\fl
E^y_{i+\frac{1}{2},j,k+\frac{1}{2}} =\frac{1}{4}\left(
	\tilde{F}^x_{i+\frac{1}{2},j,k} + 
	\tilde{F}^x_{i+\frac{1}{2},j,k+1}-
	\tilde{F}^z_{i,j,k+\frac{1}{2}} - 
	\tilde{F}^z_{i+1,j,k+\frac{1}{2}}\right) \; ,
\end{equation}
where the fluxes $\tilde{F}^i$ are those computed with the approximate
Riemann solver.

Since we are using HRSC methods, all the quantities are located at
cells centers but in equation~\eref{eq:CTdiscrete} we are effectively
evolving the magnetic field at the surfaces between the cells. The
relation between these two different values of the magnetic field is
given by a simple average
\begin{eqnarray}
\tilde{B}^x_{i,j,k} &=& \frac{1}{2} \left(
	\tilde{B}^x_{i+\frac{1}{2},j,k} + 
        \tilde{B}^x_{i-\frac{1}{2},j,k}\right) \; , \\ 
\tilde{B}^y_{i,j,k} &=& \frac{1}{2} \left(
	\tilde{B}^y_{i,j+\frac{1}{2},k} + 
        \tilde{B}^y_{i,j-\frac{1}{2},k}\right) \; , \\ 
\tilde{B}^z_{i,j,k} &=& \frac{1}{2} \left(
	\tilde{B}^z_{i,j,k+\frac{1}{2}} + 
        \tilde{B}^z_{i,j,k-\frac{1}{2}}\right) \; .
\end{eqnarray}

To demonstrate that this method guarantees that $\nabla \cdot
\vec{\tilde{B}} =0 $ and will not grow in time, we can integrate over
the volume of a numerical cell this constraint and then using the Gauss
theorem we obtain
\begin{equation}
\int_{\Delta V} \nabla \cdot \vec{\tilde{B}}\mathrm{d}V = \sum_{i=1}^{6} \int_{\Sigma_i} \vec{\tilde{B}}\cdot\mathrm{d}\vec{\Sigma} \; ,
\end{equation}
where the sum is taken over all the six faces $\Sigma_i$ that surround
the cell. Taking now the time derivative of this expression and using
eq.~\eref{eq:CTintegral} we obtain
\begin{equation}
\partial_t \int_{\Delta V} \nabla \cdot \vec{\tilde{B}}\mathrm{d}V =
-\sum_{i=1}^{6}\int_{\partial\Sigma_i} \vec{E} \cdot \vec{l} \; ,
\end{equation}
and the sum on the right-hand side gives exactly zero since the value
of $\vec{E}\cdot\vec{l}$ for the common edge of two adjacent faces has
a different sign.

\subsection{\label{sec:conversion}Primitive-variables recovery}

Because the fluxes $\mathbf{F}^i$ depend on the primitive variables
${\boldsymbol P}$ and not on the evolved conservative variables
$\mathbf{F}^0$, the values for the primitive variables need to be
recovered after each timestep and at each gridpoint. With the exception
of the magnetic field variables $B^i$, the complexity of the system of
equations to be solved prevents from an analytic solution relating the
primitive to the conservative variables through simple algebraic
relations and thus the system of equations~\eref{defD}--\eref{deftau}
needs to be solved numerically. Several methods are available for this,
the most obvious (and expensive) one consisting of solving the full set
of 5 equations given by the expressions for $(D,S_j,\tau)$ in the 5
unknowns $(\rho,v^i,\epsilon)$; we refer to this as to the 5D method.
Alternatively, and under certain conditions, it is possible to reduce the
set of equations to be solved to a couple of nonlinear equations (2D
method) or even to a single one (1D method). We review them briefly in
the following Sections but a more detailed discussion can be found in
~\cite{noble06}.

\subsubsection{\label{sec:con2prim2D}2D method}

~~~The following procedure is the same used in~\cite{anton06} and it is
an extension to full General Relativity of the method developed
in~\cite{komissarov99} in special relativity. The idea is to take the
modulus $S^2=S^jS_j$ of the momentum instead of the expression for its
three components reducing the total number of equations that one has to
solve. Using the relations~\eref{eq:b0b2} it is possible to write $S^2$
as
\begin{equation}
S^2=(Z+B^2)^2\frac{W^2-1}{W^2}-(2Z+B^2)\frac{(B^iS_i)^2}{Z^2} \; ,
\label{eq:S2}
\end{equation}
where $Z\equiv \rho h W^2$. It is also possible to rewrite the equation
for the total energy in a similar way
\begin{equation}
\tau = Z+B^2-p-\frac{B^2}{2W^2}-\frac{(B^iS_i)^2}{2Z^2}-D \; .
\label{eq:tau}
\end{equation}
Using then the definition of $D=\rho W$, eqs.~\eref{eq:S2}
and~\eref{eq:tau} form a closed system for the two unknowns $p$ and $W$,
assuming the function $h=h(\rho,p)$ is provided. When using a polytropic
EOS [i.e. eq.~\eref{poly}], the integration of the total energy equation is
not necessary (the energy density can be computed algebraically from
other quantities) and the system reduces to the numerical solution of the
equation~\eref{eq:S2}. Once the roots for $W$, $p$ and $\rho=D/W$ are
found, it is possible to compute the values of $v_i$ using the definition
of the momentum $S_i$
\begin{equation}
v_i = \frac{B_i(B^jS_j)+S_i Z}{Z(B^2+Z)} \; .
\label{eq:vel}
\end{equation}

\subsubsection{\label{sec:con2prim1D}1D method}
~~~The basic idea of this method is to consider also the gas pressure $p$
as a function of $W$ reducing the total number of equations that must be
solved numerically. When using an ideal-fluid [i.e. eq.~\eref{id_fluid}],
$Z$ can in fact be rewritten as
\begin{equation}
Z = D W + \frac{\Gamma}{\Gamma-1} p(W) W^2 \; .
\label{eq:Zp}
\end{equation}
Using equation~\eref{eq:Zp} it is possible to rewrite~\eref{eq:tau}
as a cubic equation for $p(W)$ which admits only one physical
solution. So at the end we need only to solve equation~\eref{eq:S2} for
the only unknown $W$. Having obtained $W$, we can then compute
$p=p(W)$ and the other quantities in the same way as done in the 2D
method.

\begin{table}
\caption{\label{tab:riemann}Initial conditions for the Riemann problems used to test the code.}
\begin{tabular}{|lrcccccccc|}
\cline{1-10} & & & & & & & & &\\
{\bf Test type} &{\bf State} &  $\rho$ & $p$ & $v^x$ & $v^y$  &$v^z$ & $B^x$ & $B^y$ & $B^z$ \\
\cline{1-10} & & & & & & & & &\\
{\bf Balsara Test 1} & & & & & & & & & \\
($\Gamma=2$)& {\it left}  & 1.000 & 1.0  & 0.0 & 0.0 & 0.0 & 0.5 & 1.0   & 0.0 \\
~ & {\it right} & 0.125 & 0.1  & 0.0 & 0.0 & 0.0 & 0.5 & -1.0  & 0.0 \\
\cline{1-10} & & & & & & & & &\\
{\bf Balsara Test 2} & & & & & & & & & \\
($\Gamma=5/3$)& {\it left}  & 1.0 & 30.0 & 0.0 & 0.0 & 0.0 & 5.0 & 6.0  & 6.0 \\
~ & {\it right} & 1.0 & 1.0  & 0.0 & 0.0 & 0.0 & 5.0 & 0.7  & 0.7 \\
\cline{1-10} & & & & & & & & &\\
{\bf Balsara Test 3} & & & & & & & & & \\
($\Gamma=5/3$)& {\it left}  & 1.0 & 1000.0 & 0.0 & 0.0 & 0.0 & 10.0 & 7.0  & 7.0 \\
~ & {\it right} & 1.0 & 0.1    & 0.0 & 0.0 & 0.0 & 10.0 & 0.7  & 0.7 \\
\cline{1-10} & & & & & & & & &\\
{\bf Balsara Test 4} & & & & & & & & & \\
($\Gamma=5/3$)& {\it left}  & 1.0 & 0.1  & 0.999  & 0.0 & 0.0 & 10.0 & 7.0   & 7.0 \\
~ & {\it right} & 1.0 & 0.1  & -0.999 & 0.0 & 0.0 & 10.0 & -7.0  & -7.0 \\
\cline{1-10} & & & & & & & & &\\
{\bf Balsara Test 5} & & & & & & & & & \\
($\Gamma=5/3$)& {\it left}  & 1.08 & 0.95 & 0.40  & 0.3  & 0.2 & 2.0 & 0.3   & 0.3 \\
~ & {\it right} & 1.00 & 1.0  & -0.45 & -0.2 & 0.2 & 2.0 & -0.7  & 0.5 \\
& & & & & & & & &\\
\cline{1-10}
\end{tabular}
\end{table}

\subsection{Atmosphere treatment}
\label{sec:atmosphere}

As already done in the \texttt{Whisky} code and in other full GRMHD
codes~\cite{shibata05,duez05} we avoid the presence of vacuum regions
in our domain by imposing a floor value to the rest-mass density. This
is necessary because the routines that recover the primitive variables
from the conservative ones may fail to find a physical solution if the
rest-mass density is too small. The floor value used for the tests
reported here is
$\rho_{\mathrm{atm}}=10^{-7}\times\mathrm{max}(\rho_0)$, with
$\rho_0$ being the value of the rest-mass density at $t=0$, but a
floor which is two orders of magnitude smaller works equally well. In
practice, for all of the numerical cells at which $\rho \leq
\rho_{\mathrm{atm}}$, we simply set $\rho=\rho_{\mathrm{atm}}$,
$v^j=0$, and do not modify the magnetic field. This is different from
what done in other codes (\textit{e.g.,}~\cite{shibata05,duez05}),
which set to zero the initial value of the magnetic field in the low
density regions.

\subsection{\label{excision}Excision}

Many interesting astrophysical scenarios involve the presence of black
holes and so of regions of spacetime where singularities are
present. These regions are causally disconnected from the rest of the
physical domain and the values of the fields inside should not affect
the zone outside the event horizon. This is not true in numerical
codes where it can happen that some information from inside the event
horizon is used to compute the values of the variables outside. In
order to avoid this, excision algorithms were developed in general
relativistic hydrodynamics and they are based on the use of some kind
of boundary condition applied to the boundary between the excised
zone, where the equations are no more solved, and the domain outside.
As already done in the \texttt{Whisky} code we apply a zeroth-order
extrapolation to all the variables at the boundary, i.e. a simple copy
of the MHD variables across the excision boundary. A different method,
based on the use of a linear extrapolation, has instead been
implemented in~\cite{duez05} and although it can lead to improved
accuracy for smooth flows (and especially when the MHD variables
change rapidly near the excision boundary), it also leads to
significantly incorrect results when shocks are present (see
Sect.~\ref{sec:excisiontest}). Great care must therefore be paid at
the properties of the flow near the excision boundary and the code
presently includes both algorithms.

It is important also to note that other methods, not based on excision
techniques, are being developed to improve the stability of numerical
codes when black hole are present in the domain. One of these
approaches is based on the use of a Kreiss-Oliger dissipation for the
field variables inside the apparent horizon~\cite{baiotti06} and it
can be straightforwardly extended also to the MHD case.

\subsection{Mesh Refinement}    
\label{sec:mesh_refinement}

The developments made in \texttt{Whisky} for handling non-uniform
grids have been extended also to \texttt{WhiskyMHD} which can
therefore use a ``box-in-box'' mesh refinement
strategy~\cite{Schnetter-etal-03b}. This allows to reduces the
influence of inaccurate boundary conditions at the outer boundaries
and for the wave-zone to be included in the computational domain.  In
practice, we have adopted a Berger-Oliger prescription for the
refinement of meshes on different levels~\cite{Berger84} and used the
numerical infrastructure described in~\cite{Schnetter-etal-03b}, \textit{i.e.,}
the {\tt Carpet} mesh refinement driver for {\tt Cactus} (see
\cite{carpetweb} for details). In addition to this, a simplified form
of adaptivity is also in place and which allows for new refined levels
to be added at predefined positions during the evolution or for
refinement boxes to be moved across the domain to follow, for
instance, regions where higher resolution is needed.

\begin{figure}
  \begin{center}
  \includegraphics[width=0.49\textwidth]{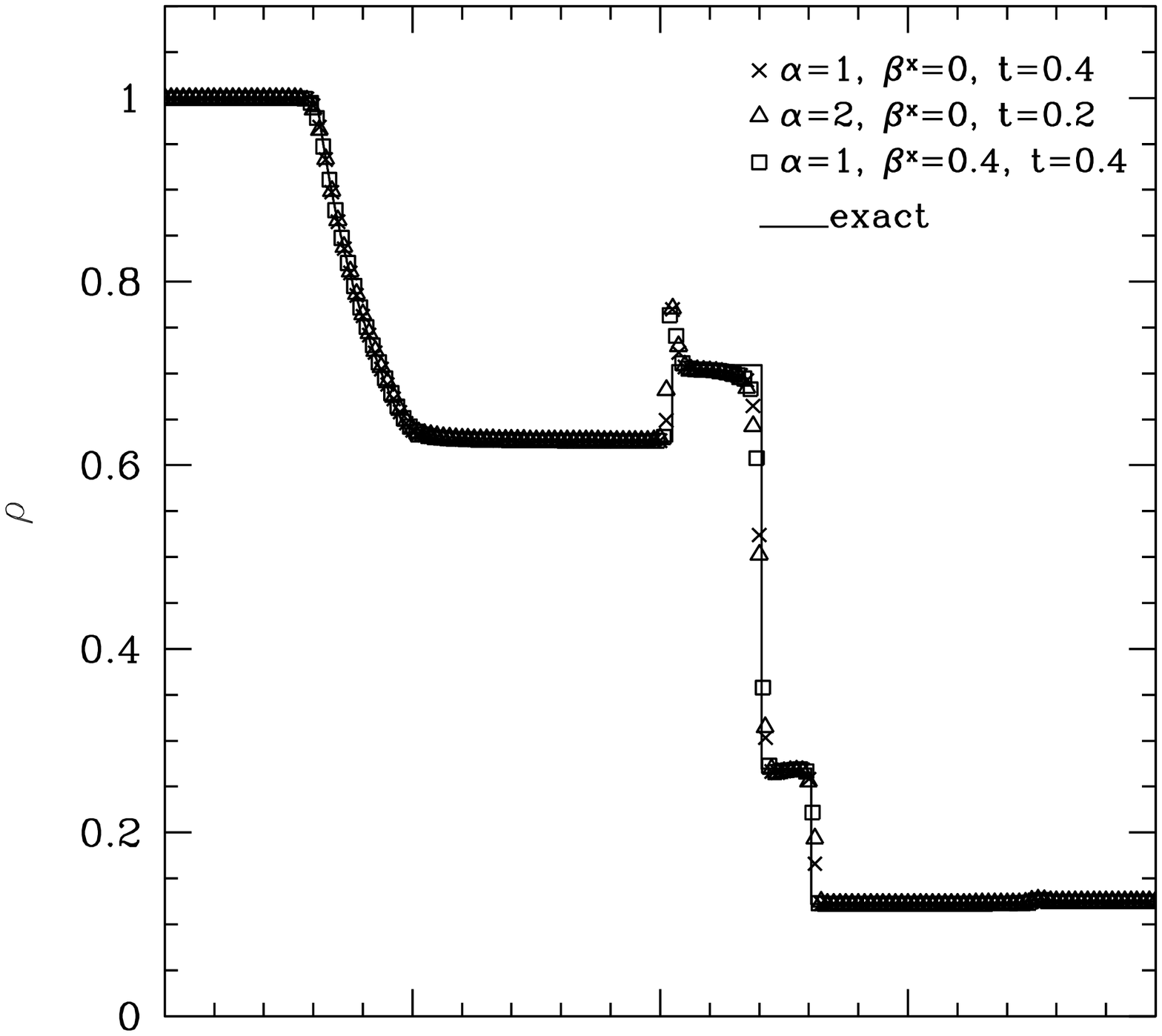}
  \hfill
  \includegraphics[width=0.49\textwidth]{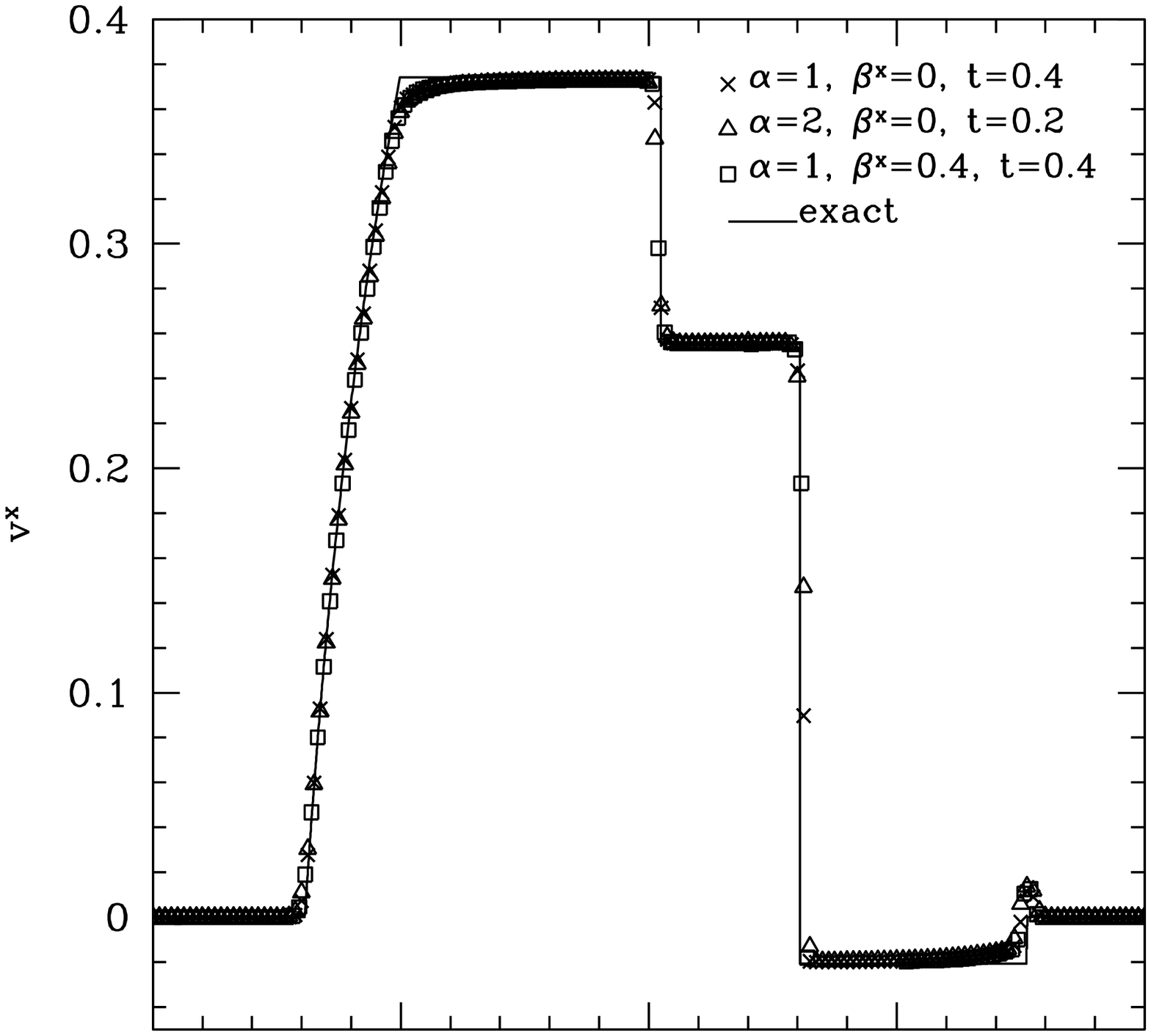}
  \vskip -1.0cm
  \includegraphics[width=0.49\textwidth]{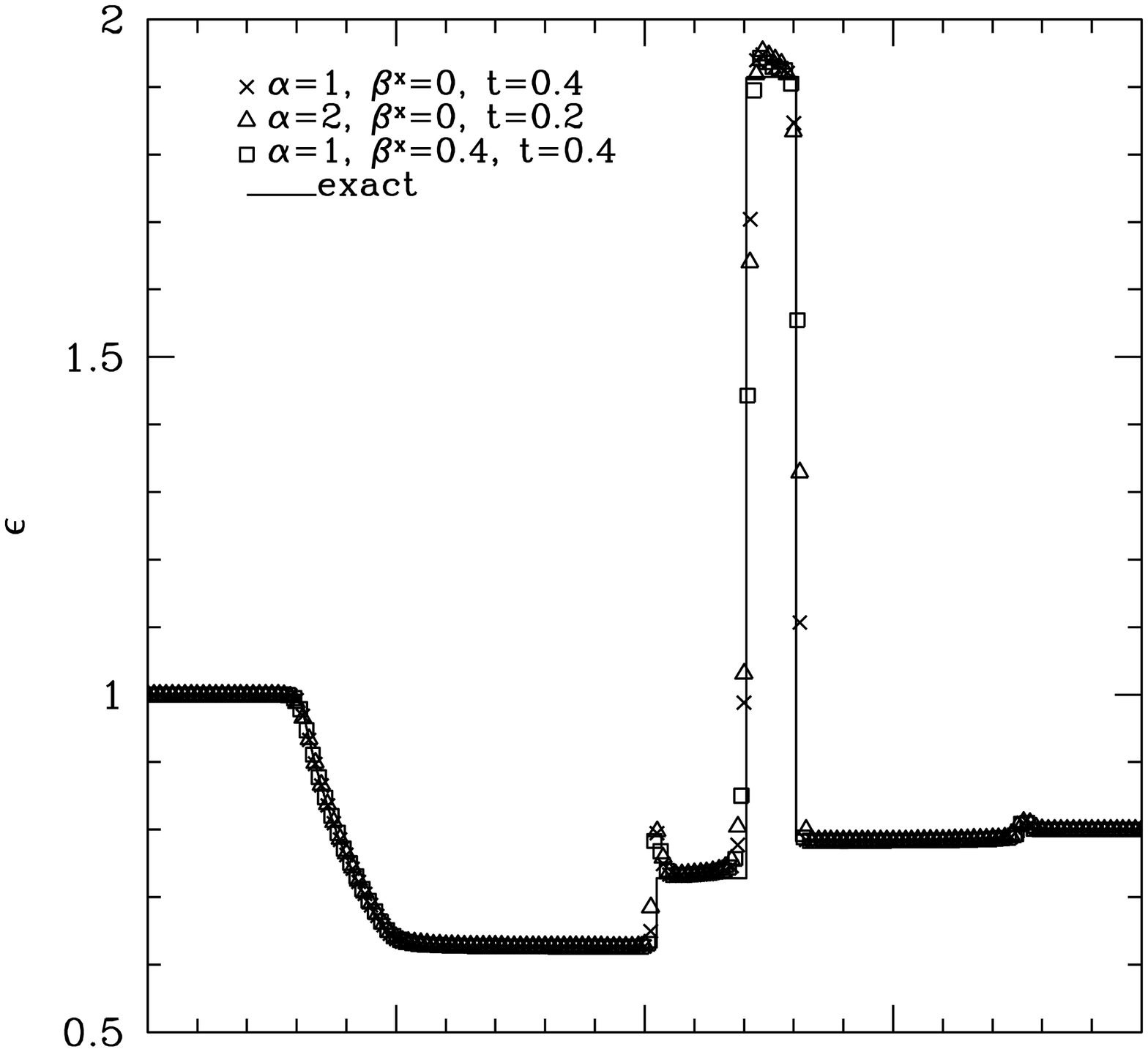}
  \hfill
  \includegraphics[width=0.49\textwidth]{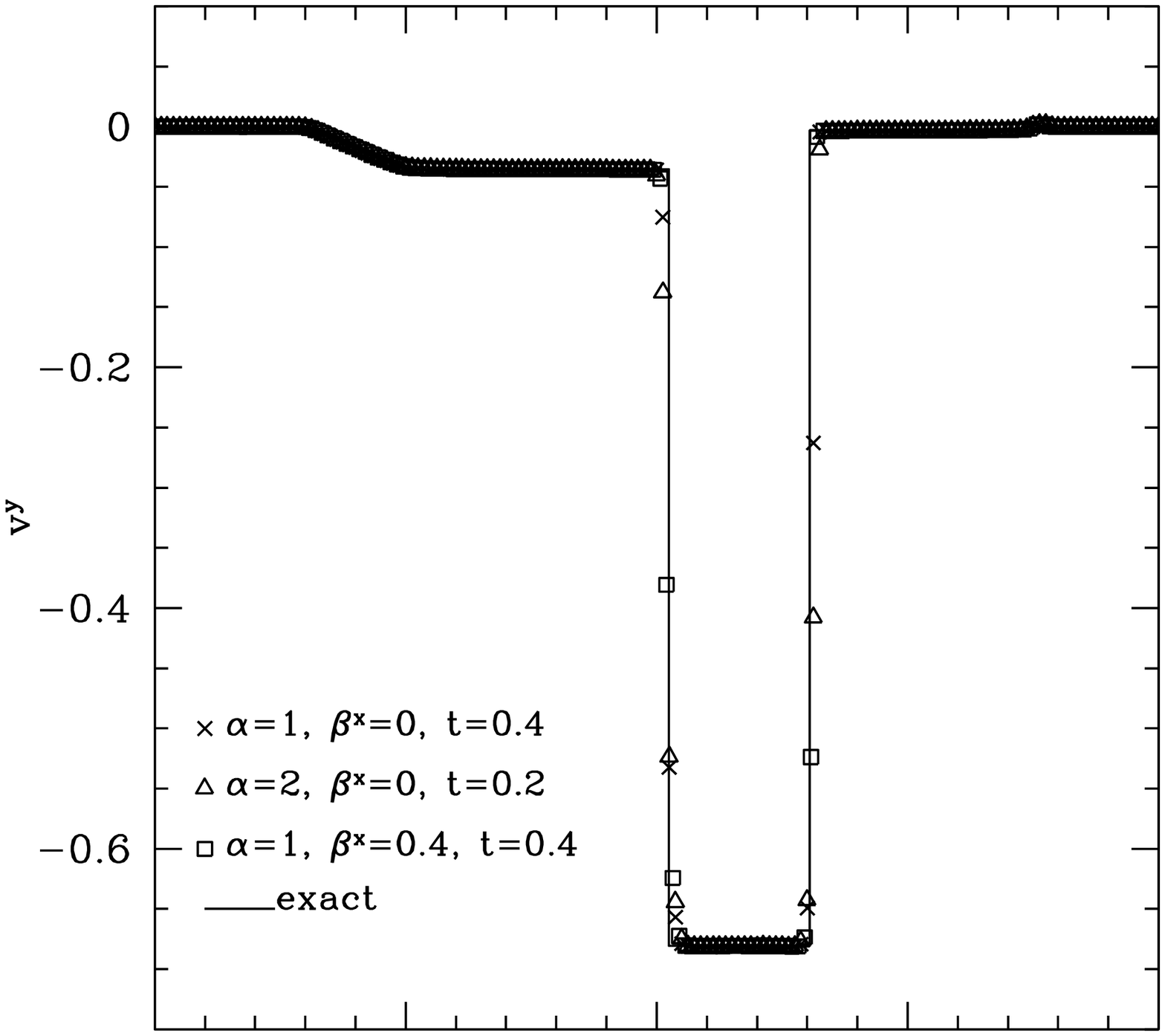}
  \vskip -1.0cm
  \includegraphics[width=0.49\textwidth]{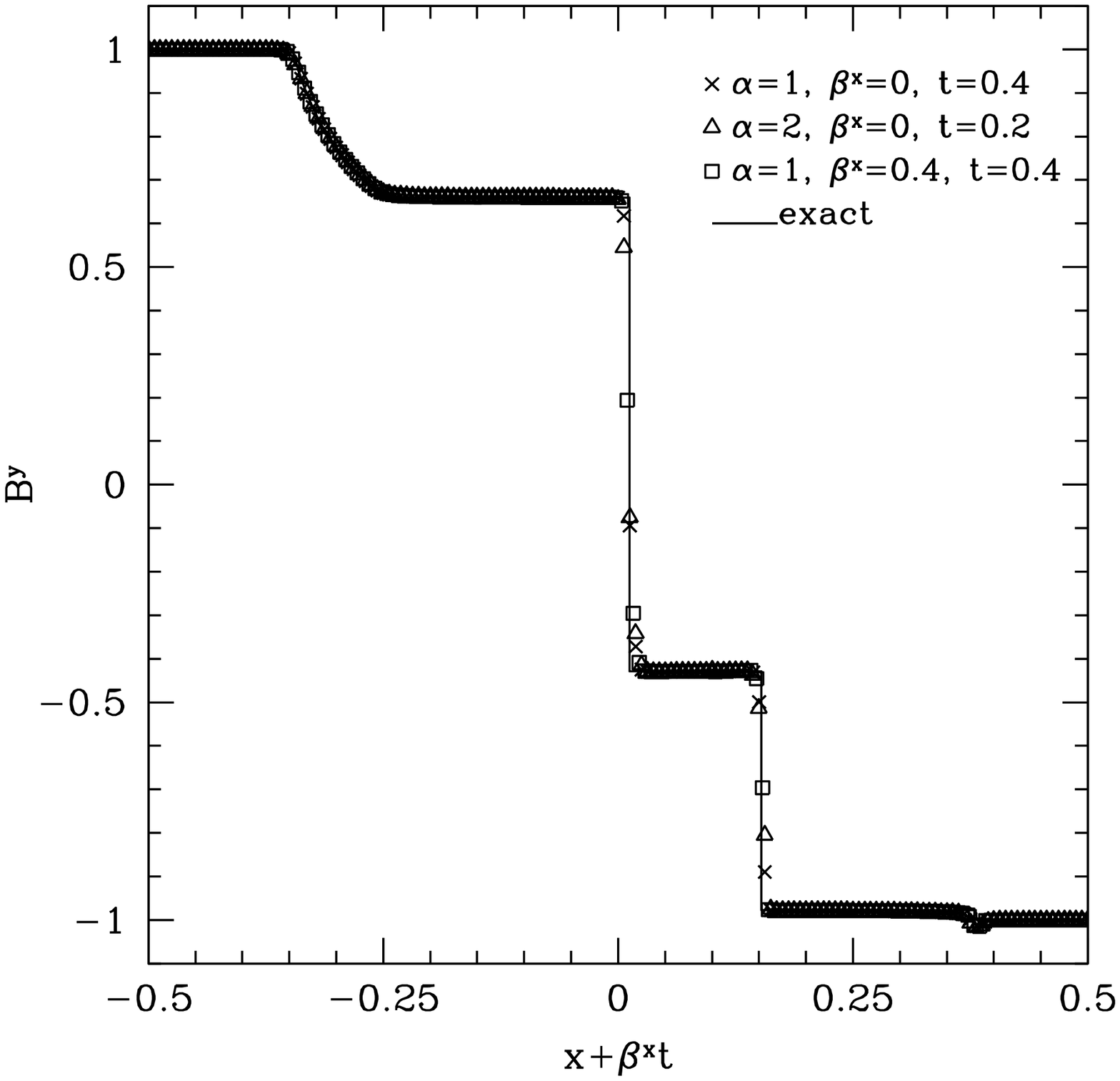}
  \hfill
  \includegraphics[width=0.49\textwidth]{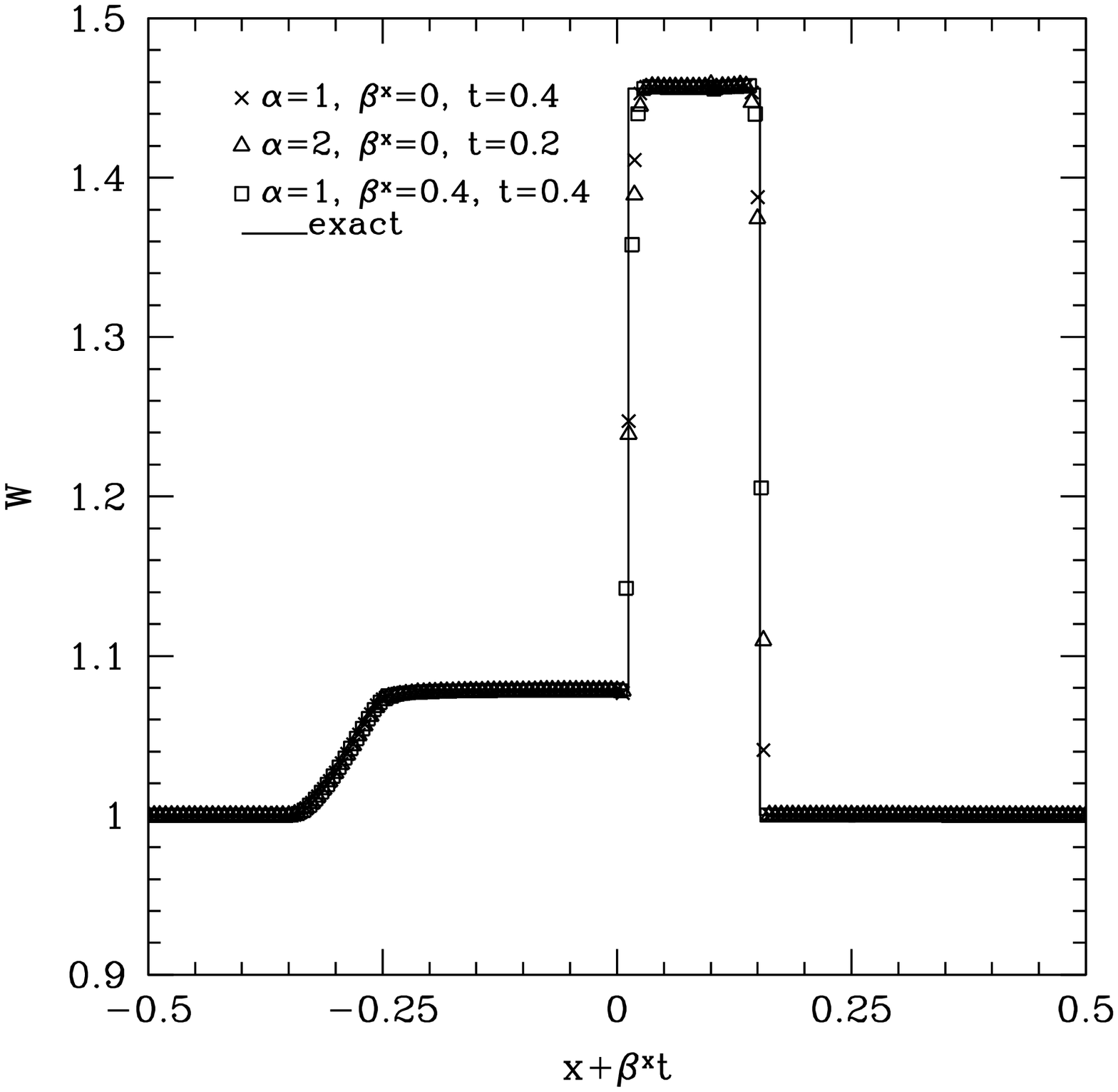}
  \end{center}
  \caption{\label{fig_whiskymhd:balsara1}Numerical solution of the
    test number 1 of Balsara with different values for the lapse
    $\alpha$ and the shift $\beta^x$. The solid line represents the
    exact solution, the crosses the numerical one at time $t=0.4$, the
    open triangle at time $t=0.2$ but with $\alpha=2$ and the open
    squares at $t=0.4$ but with $\beta^x=0.4$, in this last case the
    solution is shifted on the $x$-axis by the amount $\beta^xt$. Note
    that only 160 of the 1600 data points used in the numerical
    solution are shown.}
\end{figure}
\begin{figure}
  \begin{center}
    \includegraphics[width=0.4\textwidth]{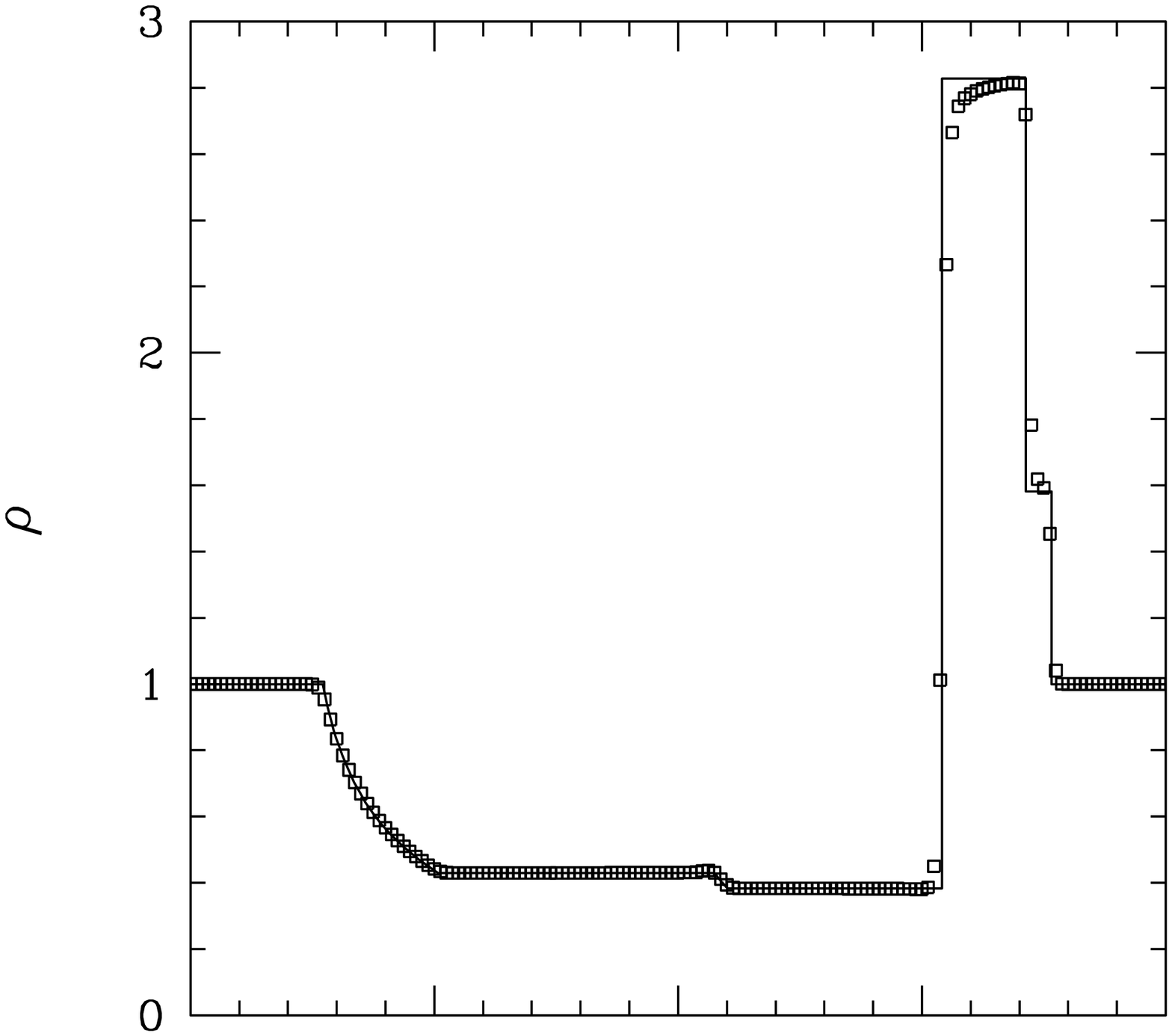}
    \hfill
    \includegraphics[width=0.4\textwidth]{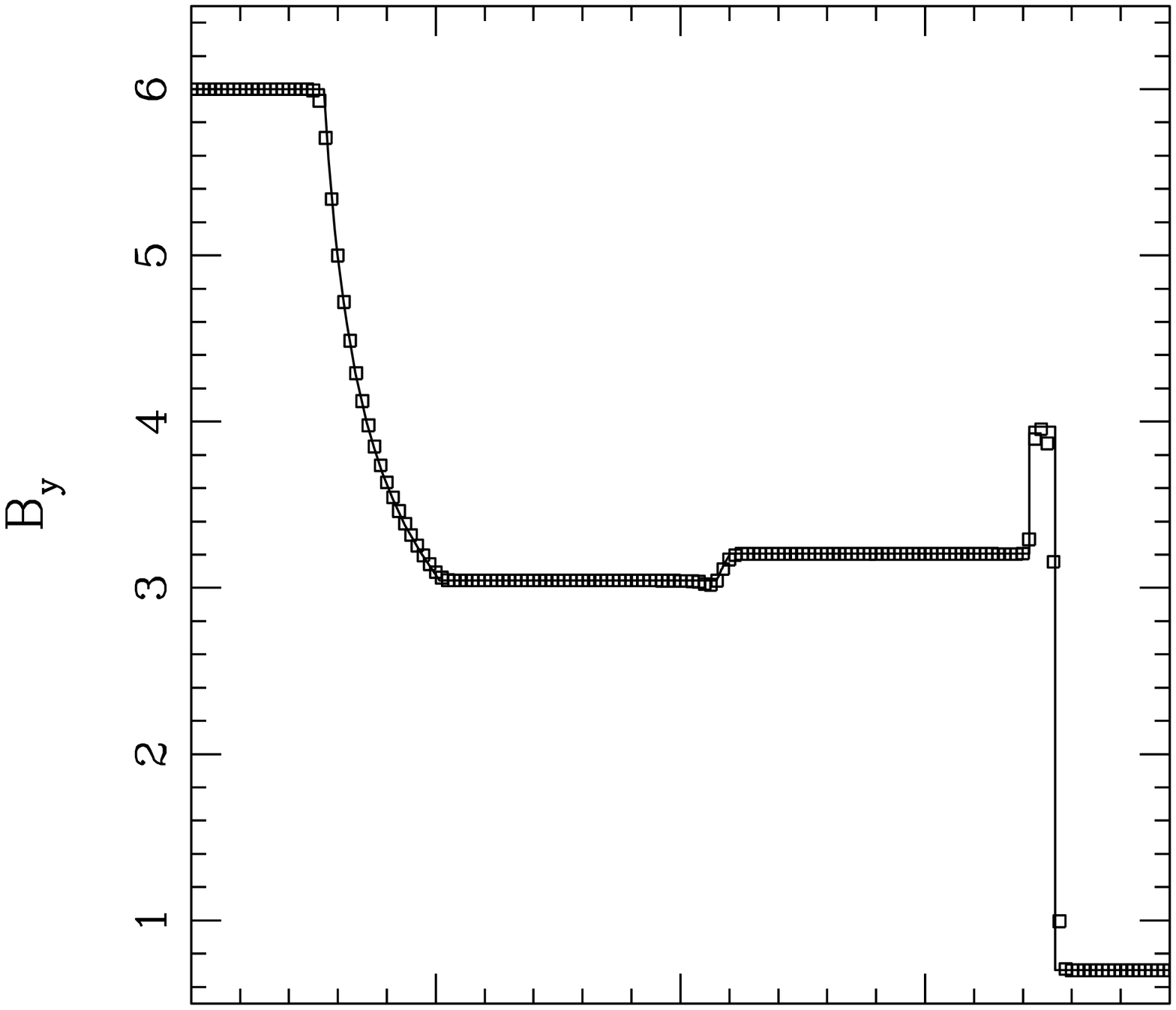}
    \vskip -0.9cm
    \includegraphics[width=0.4\textwidth]{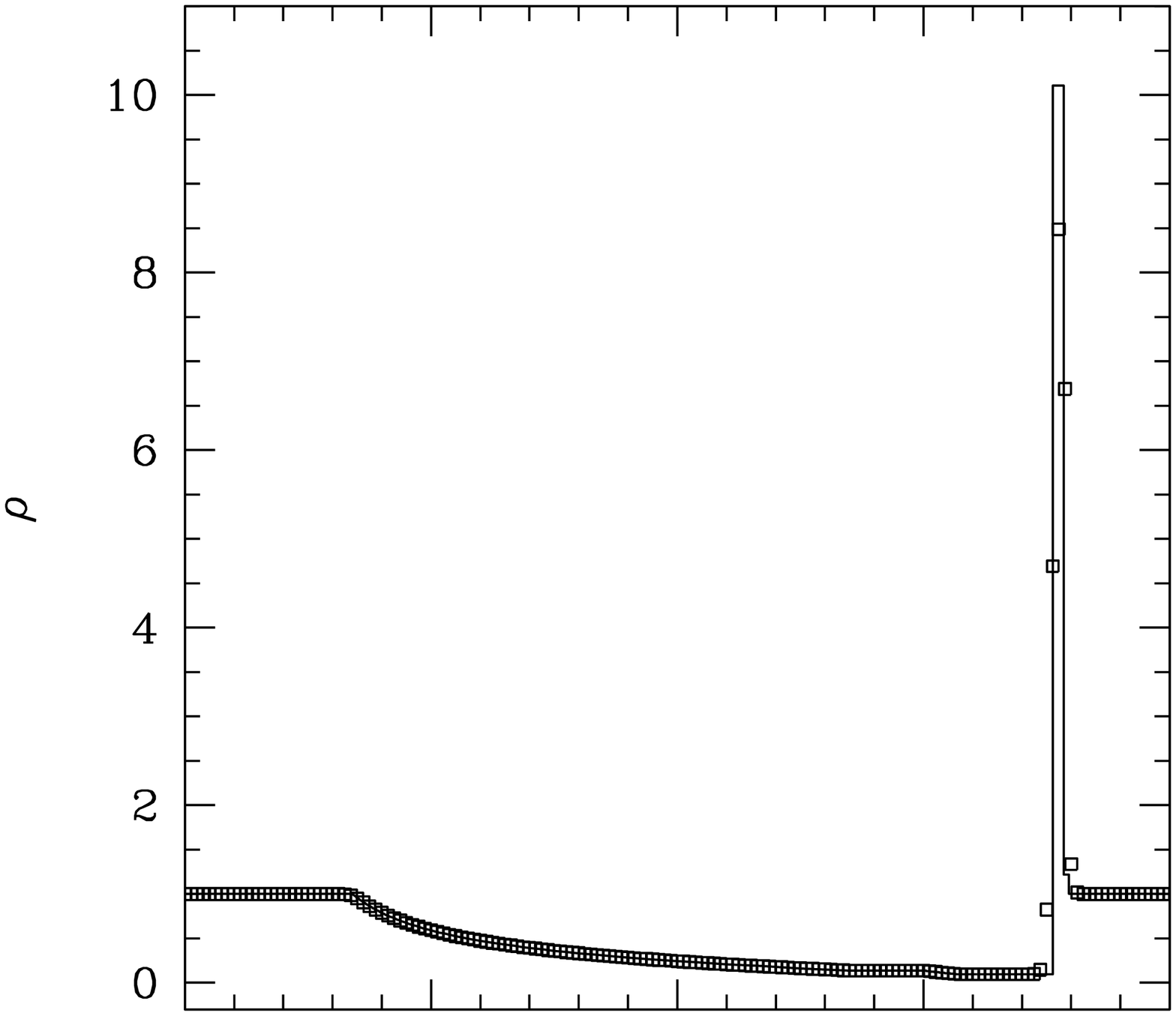}
    \hfill
    \includegraphics[width=0.4\textwidth]{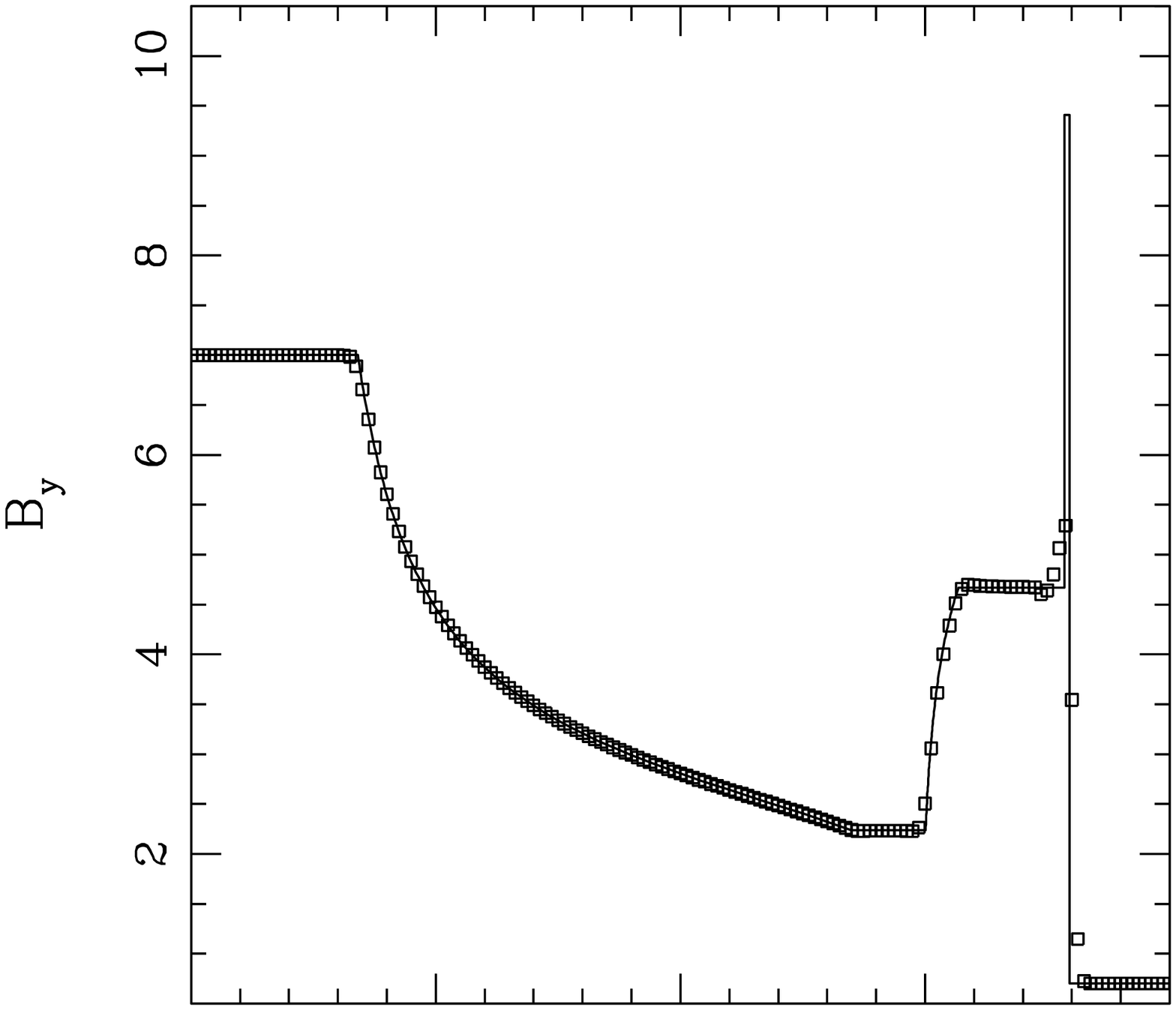}
    \vskip -0.9cm
    \includegraphics[width=0.4\textwidth]{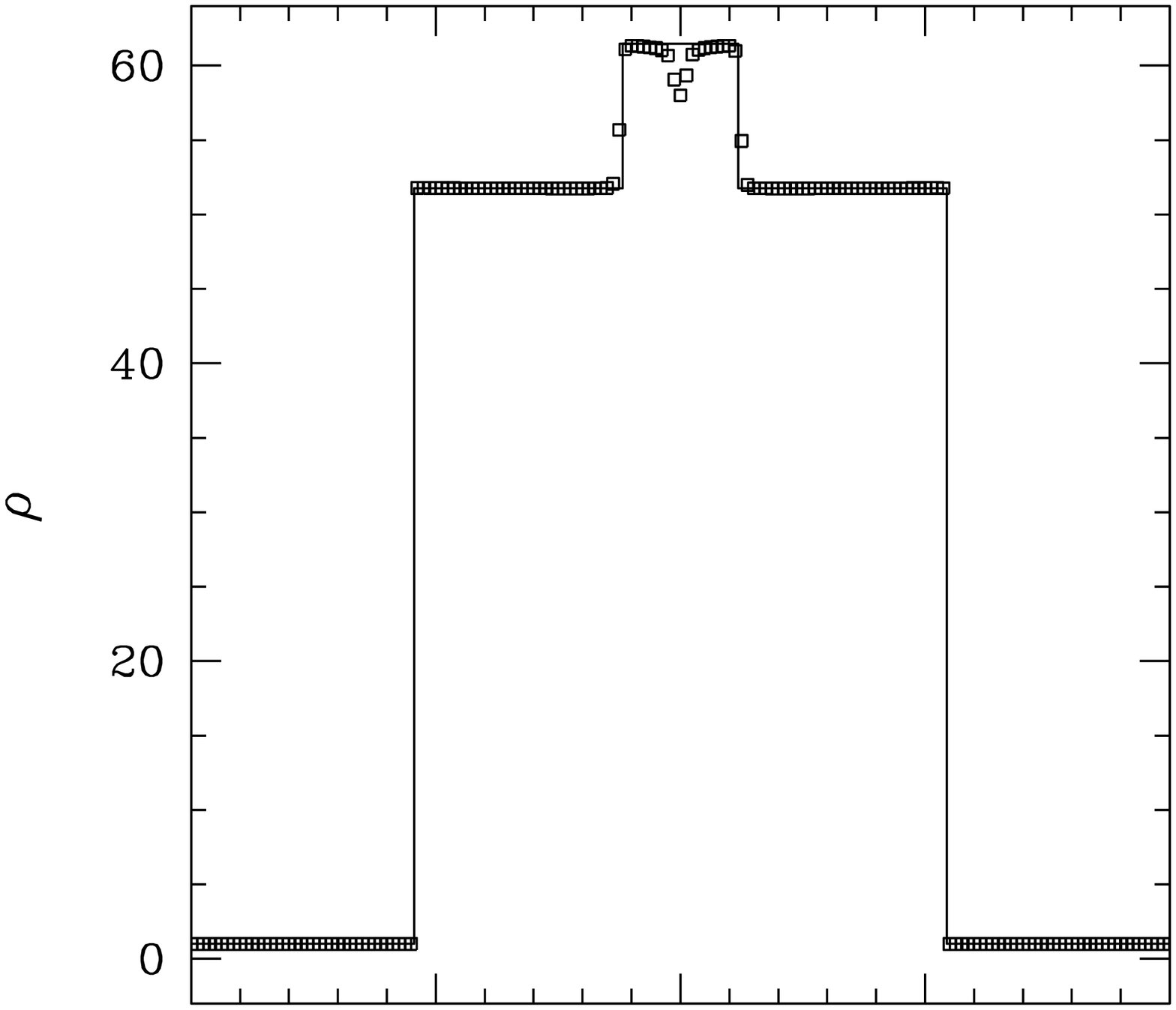}
    \hfill
    \includegraphics[width=0.4\textwidth]{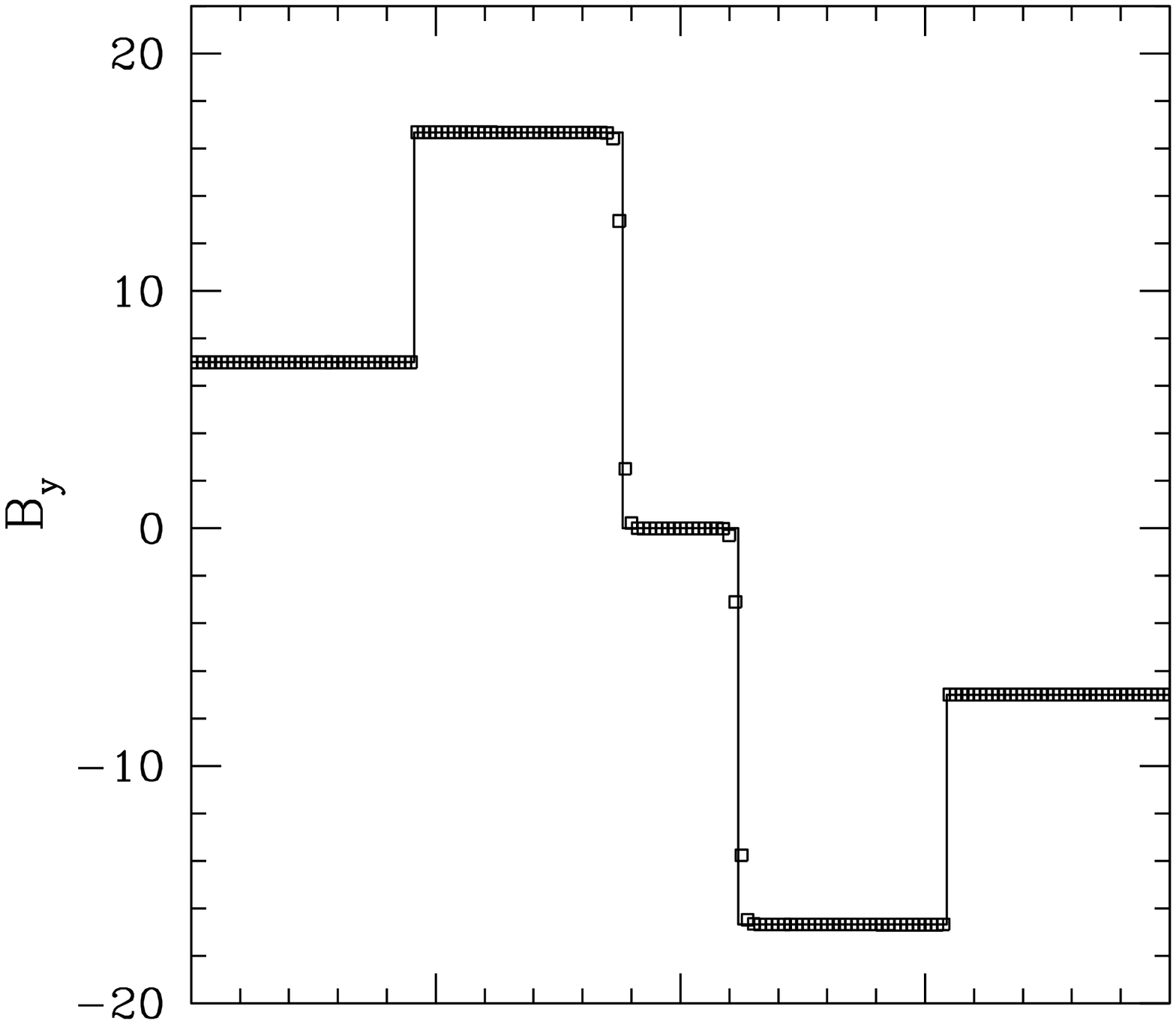}
    \vskip -0.9cm
    \includegraphics[width=0.4\textwidth]{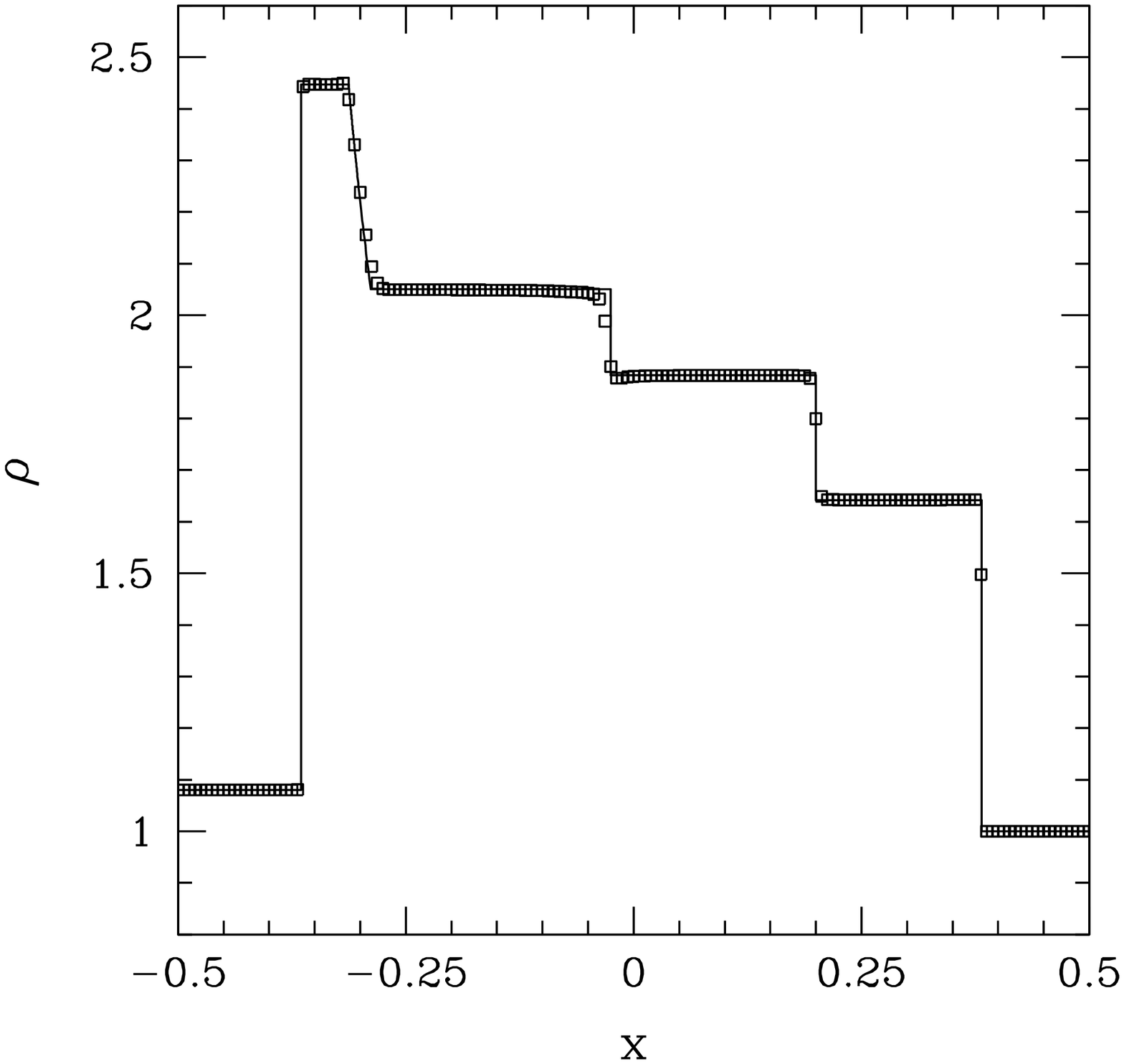}
    \hfill
    \includegraphics[width=0.4\textwidth]{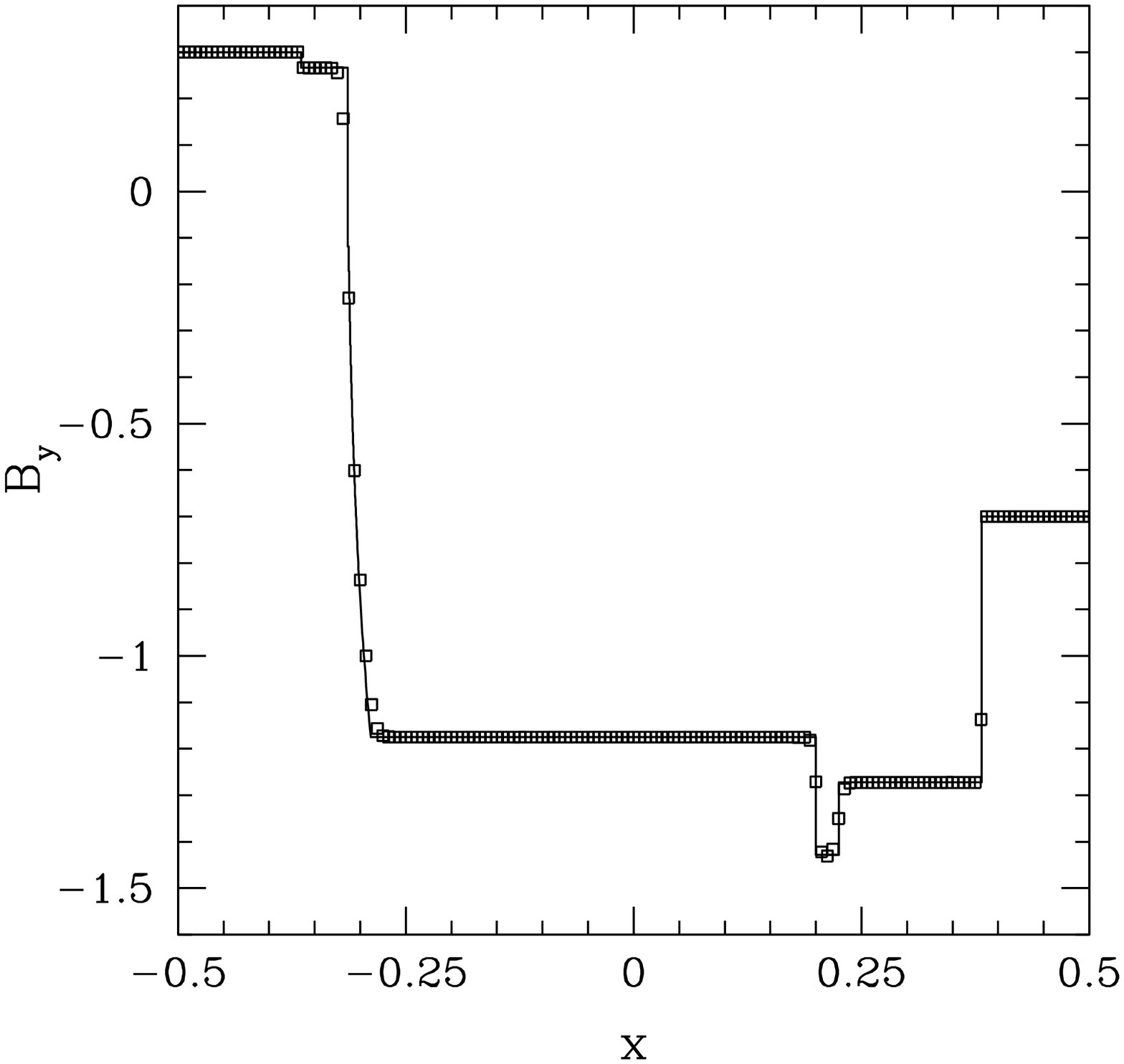}
  \end{center}
  \caption{\label{fig_whiskymhd:balsara2345} Numerical solution of the
    tests number 2 (first row), 3 (second row), 4 (third row) and 5
    (fourth row) of Balsara. The first 3 are computed at a time
    $t=0.4$ while the test number 5 is computed at $t=0.55$. The solid
    line represent the exact solution while the open squares the
    numerical one. Note that only 160 of the 1600 data points used in
    the numerical solution are shown.}
\end{figure}

\section{\label{sec:tests}Tests}

Code-testing represents an important aspect of the development of any
newly developed and multidimensional code because it validates that
all of the algorithms are implemented correctly and represent a
faithful and discretized representation of the continuum equations
they are solving. In what follows we report the results for a series
of testbeds ranging from the solution of relativistic Riemann problems
in flat spacetime, over to the stationary accretion onto a
Schwarzschild black hole and up to the evolution of isolated and
oscillating magnetized stars. We note that in the tests involving a
polytropic EOS the recovery of the primitive variables has been made
using the ``2D-method'' (see Sect.~\ref{sec:con2prim2D}), while a
``1D-method'' has been used when adopting an ideal-fluid EOS (see
Sect.~\ref{sec:con2prim1D}).


\subsection{\label{sec:balsara}Riemann problems}

As customary in the testing of hydrodynamics and magnetohydrodynamics
codes we have first validated \texttt{WhiskyMHD} against a set of Riemann
problems in a Minkowski spacetime following the series of initial
conditions proposed by Balsara~\cite{balsara01b}. All these tests were
run on a grid of unit length with 1600 grid points with the initial
discontinuity located at the center of the grid. An ideal equation of
state with $\Gamma=5/3$ was used with the exception of the first test
with $\Gamma=2$ and the initial conditions for all the tests are reported
in Table~\ref{tab:riemann}.

In all of the tests presented here the numerical solution for the
different MHD variables has been compared with the exact one
computed with the exact Riemann solver discussed
in~\cite{giacomazzo06}. This represents an important difference with
what done in the past by similar codes as it allows, for the first
time, for a quantitative assessment of the code's ability to evolve
correctly all the different waves that can form in relativistic
MHD. In figures~\ref{fig_whiskymhd:balsara1}
and~\ref{fig_whiskymhd:balsara2345} the exact solution is represented
with a solid line, while the numerical one with different symbols.

In~\fref{fig_whiskymhd:balsara1}, in particular, we show the
comparison between the numerical and the exact solution at $t=0.4$ for
several MHD variables as computed for the relativistic analogue of the
classical Brio-Wu shock tube problem~\cite{brio88,vanputten93}. The
initial discontinuity develops a left-going fast rarefaction, a
left-going slow compound wave, a contact discontinuity, a right-going
slow shock and fast rarefaction. Note that besides presenting the
solution with $\alpha=1$ and $\beta^x=0$, we have exploited the
freedom in choosing these gauges and validated the code also for less
trivial values of the lapse and shift~\cite{anton06}. More
specifically, shown with different symbols
in~\fref{fig_whiskymhd:balsara1} are the numerical solutions with
$\alpha=2$ at time $t=0.2$ and with $\beta^x=0.4$ after the latter has
been shifted in space by $\beta^x t$. Clearly, all the symbols overlap
extremely well, coinciding with the exact solution also in the
presence of strong discontinuities. Note that only 160 of the 1600
data points used in the simulation are shown and that the difference
between the numerical solution and the exact one at the compound wave
is due to the fact that, by construction, our exact solver assumes
compound waves never form. We have indeed adopted the same standpoint
of Ryu and Jones~\cite{Ryu95} in the development of their exact
Riemann solver in nonrelativistic magnetohydrodynamics. We also remark
that it is not yet clear whether compound waves have to be considered
acceptable physical solutions of the ideal MHD equations and a debate
on this is still ongoing (see, for instance,
\cite{falle01,torrilhon03a,torrilhon03b,torrilhon04}).

Similarly, shown in~\fref{fig_whiskymhd:balsara2345} are the
comparisons between the numerical solution for the rest-mass density
and the $y$-component of the magnetic field for the tests number 2
(first row), 3 (second row), 4 (third row) and 5 (fourth row) of
Balsara. The first 3 are computed at a time $t=0.4$ while the test
number 5 is computed at $t=0.55$. Clearly, our code is able to resolve
all the different waves present in MHD, showing a very good agreement
with the exact solution. Other Riemann problems have been carried out
in different directions (either along coordinate axes or along main
diagonals) and they all provide the same level of accuracy discussed
in figures~\ref{fig_whiskymhd:balsara1}
and~\ref{fig_whiskymhd:balsara2345}.

\begin{figure}
  \begin{center}
  \includegraphics[width=0.47\textwidth]{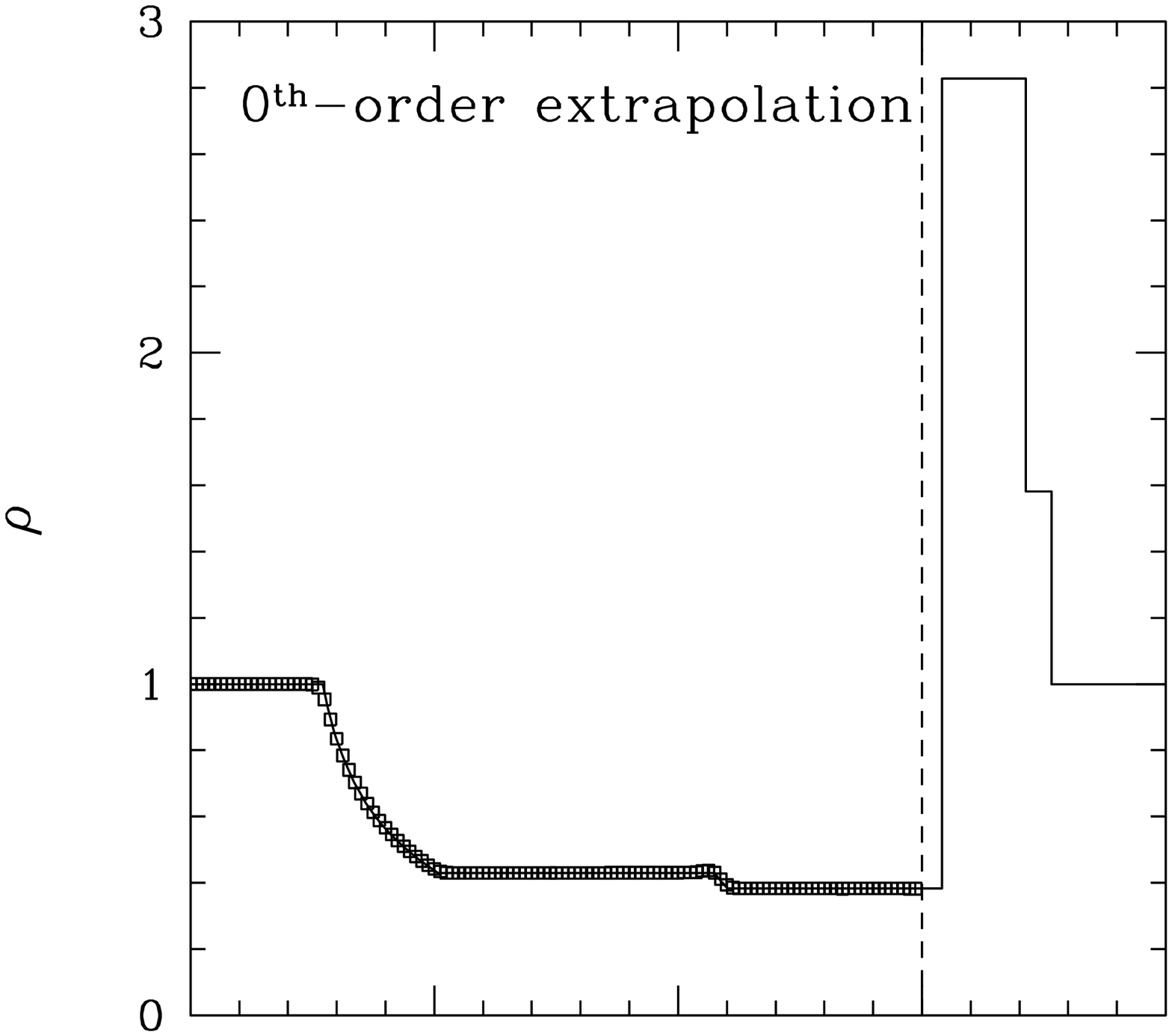}
  \hfill
  \includegraphics[width=0.47\textwidth]{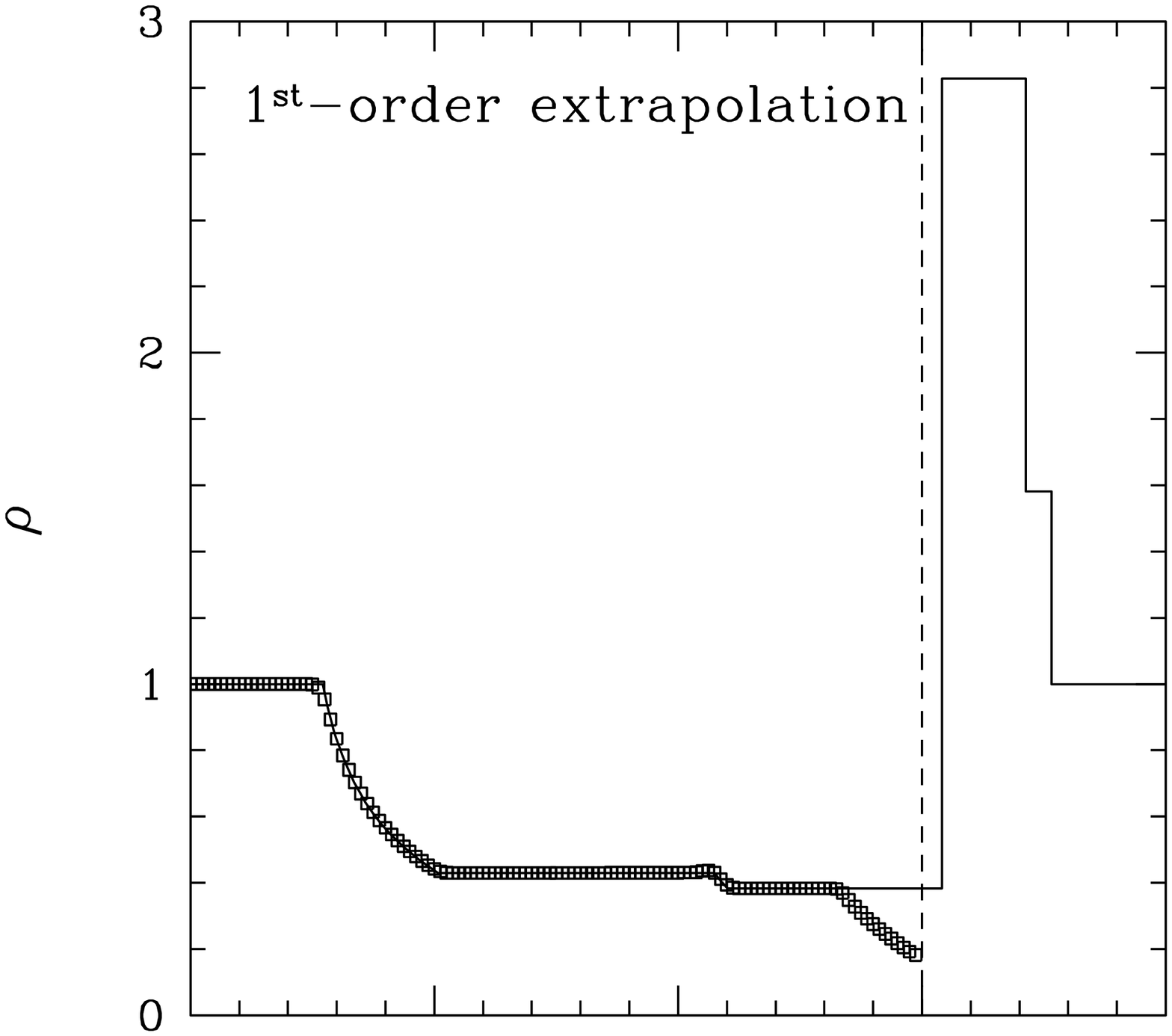}
  \vskip -1.0cm
  \includegraphics[width=0.47\textwidth]{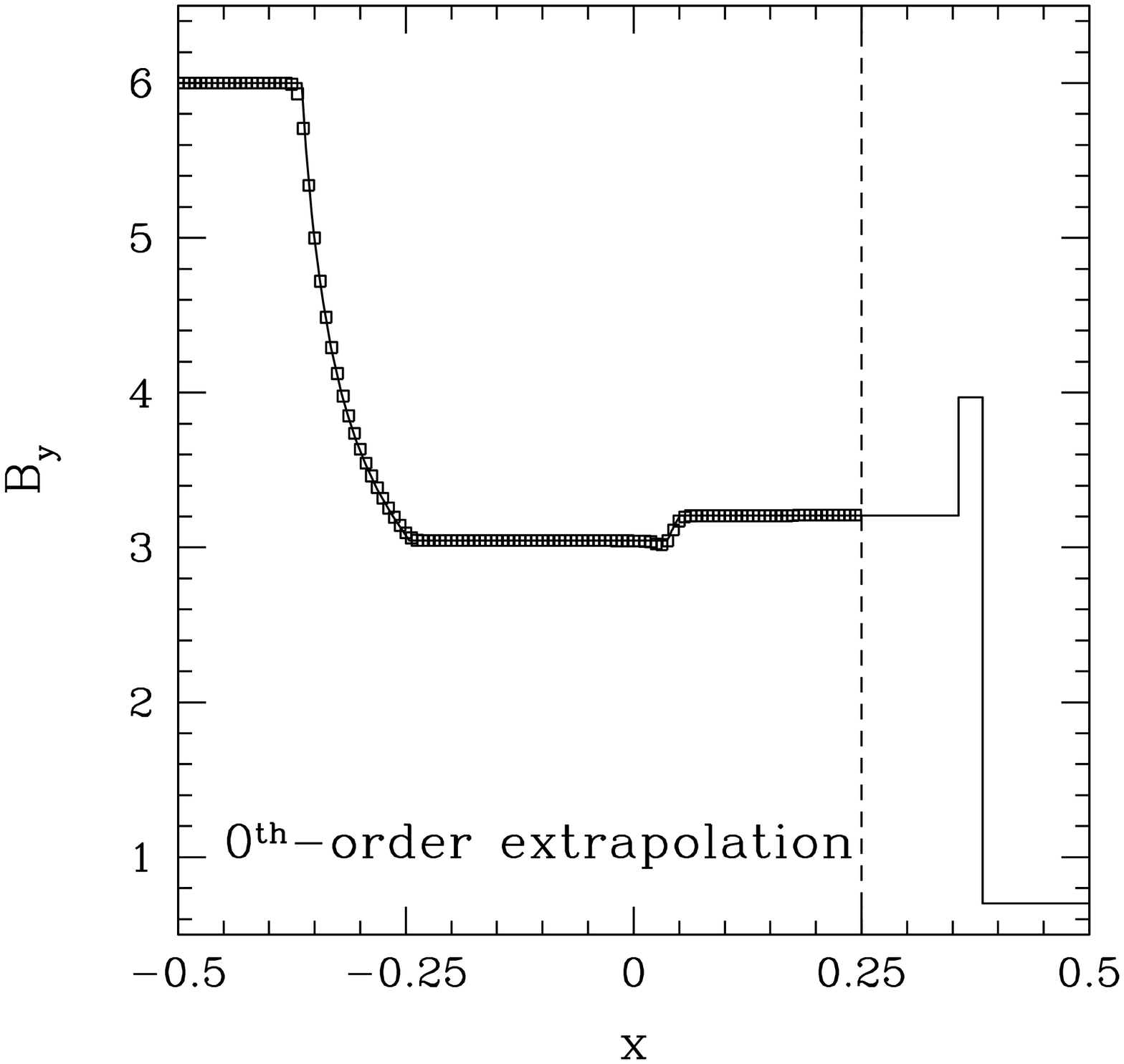}
  \hfill
  \includegraphics[width=0.47\textwidth]{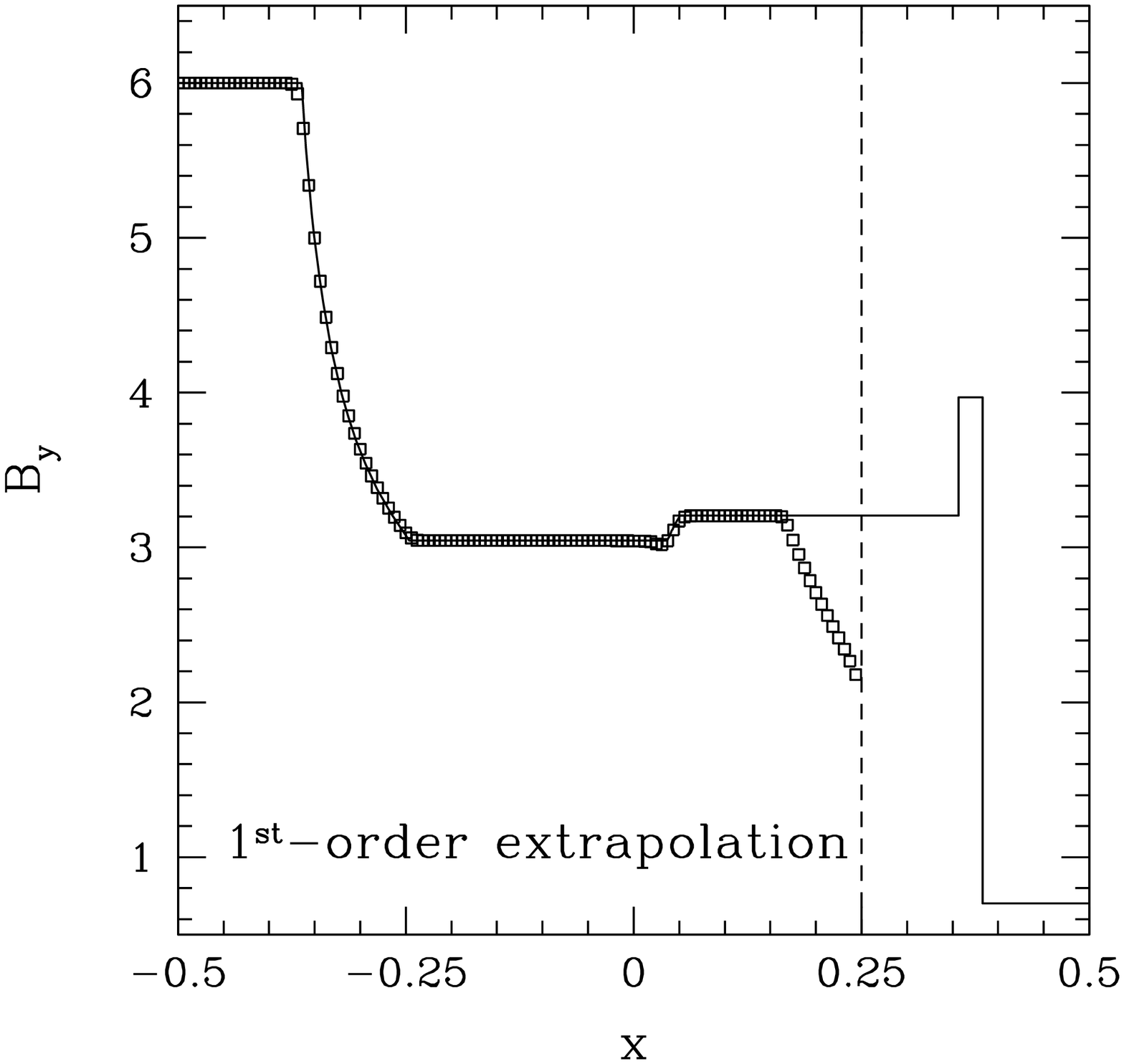}
  \end{center}
  \caption{\label{fig_whiskymhd:balsara2_excision} Numerical solution
    of the test number 2 of Balsara at a time $t=0.4$ with an excision
    boundary (dashed vertical line) located at $x=0.25$; the region at
    the right of this boundary is not evolved. In the two left panels
    a zeroth-order extrapolation, {\it i.e.} a simple copy, was used,
    while in the two right panels the values of the different
    variables at the excision boundary were obtained with a linear
    extrapolation. The solid line represents the exact solution while
    the open squares the numerical one. The solution is composed of
    two left-going fast and slow rarefactions, of a contact
    discontinuity and of two right-going fast and slow shocks. Only
    160 of the 1600 data points used in the numerical solution are
    shown.}
\end{figure}

\subsection{\label{sec:excisiontest}Excision tests on a flat background}

We next show the code's ability to accurately evolve shocks also when
an excised region is present in the domain. To this scope, we have
used the test number 2 of Balsara excising the region $x^i
\in[0.25,0.5]$ and using the zeroth-order extrapolation scheme. In
this case the fast and slow shocks moving to the right go inside the
excised region and the solution outside is not affected. This is shown
in the left panels of~\fref{fig_whiskymhd:balsara2_excision} which
report with small squares the numerical solution for $\rho$ and $B_y$
of the test number 2 of Balsara with an excision boundary (dashed
vertical line) located at $x=0.25$. The data refers to time $t=0.4$,
when the right-going waves have already gone through the excision
boundary as indicated by the exact solution (continuous line).

As a comparison, and to underline its incorrectness in the case of
non-smooth flows, we have also considered boundary conditions
involving a linear extrapolation of the MHD variables across of the
excision boundary as suggested in ~\cite{duez05}. This is illustrated
in the right panels of~\fref{fig_whiskymhd:balsara2_excision} which
are the same as the left ones but for the different boundary condition
at the excision boundary. Clearly, in this case the solution outside
the excision region is badly affected and a left-going wave is
produced which rapidly spoils the solution. Because this happens only
when the discontinuity crosses the excision boundary, it is clear that
a linear extrapolation is not adequate in this case as it provides an
incorrect information on the causal structure of the flow near the
boundary. As we will discuss in the following Section, however, a
linear extrapolation remains a good, and sometimes preferable, choice
in the case of smooth flows.

\subsection{\label{sec:michel}Magnetized spherical accretion}

This second test proves the ability of the code to evolve accurately
stationary accretion solution in a curved but fixed spacetime. In
particular, we consider the spherical accretion of a perfect fluid
with a radial magnetic field onto a Schwarzschild black hole (this is
sometimes referred to as a relativistic Bondi flow). The solution to
this problem is already known for the unmagnetized case, but it is
simple to show that its form is not affected if a radial magnetic
field is added~\cite{devilliers03}. The initial setup for this test is
the same used in~\cite{devilliers03,harm,duez05,anton06} and consists
of a perfect fluid obeying a polytropic EOS with $\Gamma=4/3$. The
critical radius of the solution is located at $r_c=8M$ and the
rest-mass density at $r_c$ is $\rho_c=6.25\times 10^{-2}$. These
parameters are sufficient to provide the full description of the
accretion onto a solar mass Schwarzschild black hole as described
in~\cite{michel72}. We solve the problem on a Cartesian grid going
from $x^i=0$ to $x^i=11M$.

To avoid problems at the horizon, located at $r=2M$, the metric is
written in terms of ingoing Eddington-Finkelstein coordinates. The
excision boundary has the shape of a cubical box of length $M$ so that
the domain $[0,M]\times[0,M]\times[0,M]$ is excluded from the
evolution. Furthermore, as a boundary condition across the excised
cube we have considered both a zero-th and a first-order
extrapolation, finding the latter to yield sligthly more accurate
results (e.g., the overall error is smaller of $\approx 3\%$ for a
test case with $b^2/\rho=25$, where $b^2/\rho$ is the dimensionless
magnetic field strength as measured at $r=2M$). As discussed earlier,
this is indeed to be expected for smooth flows as the ones considered
here for the relativistic Bondi flow.

\begin{figure}
  \begin{center}
  \includegraphics[width=0.47\textwidth]{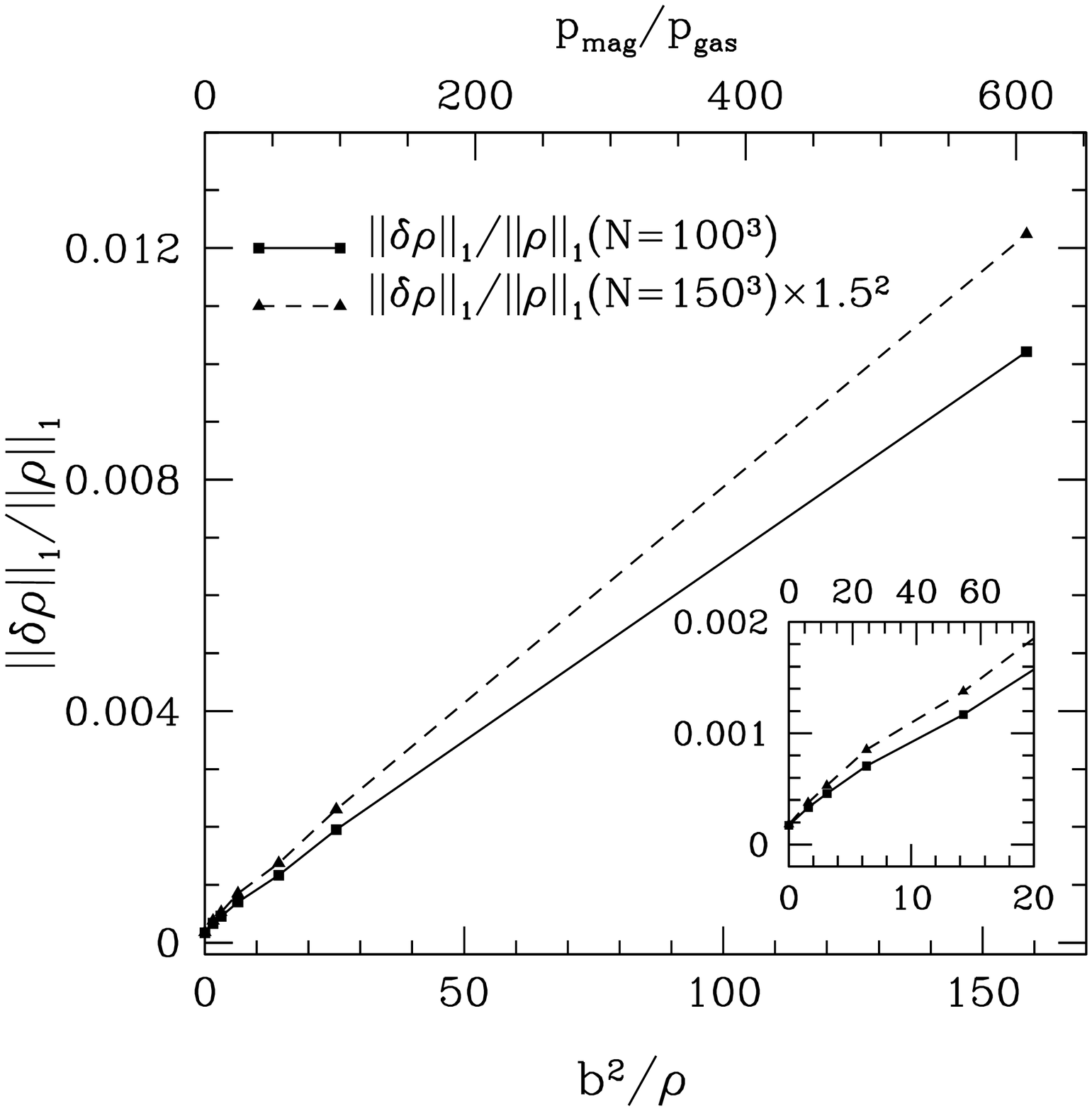}
  \hfill
  \includegraphics[width=0.47\textwidth]{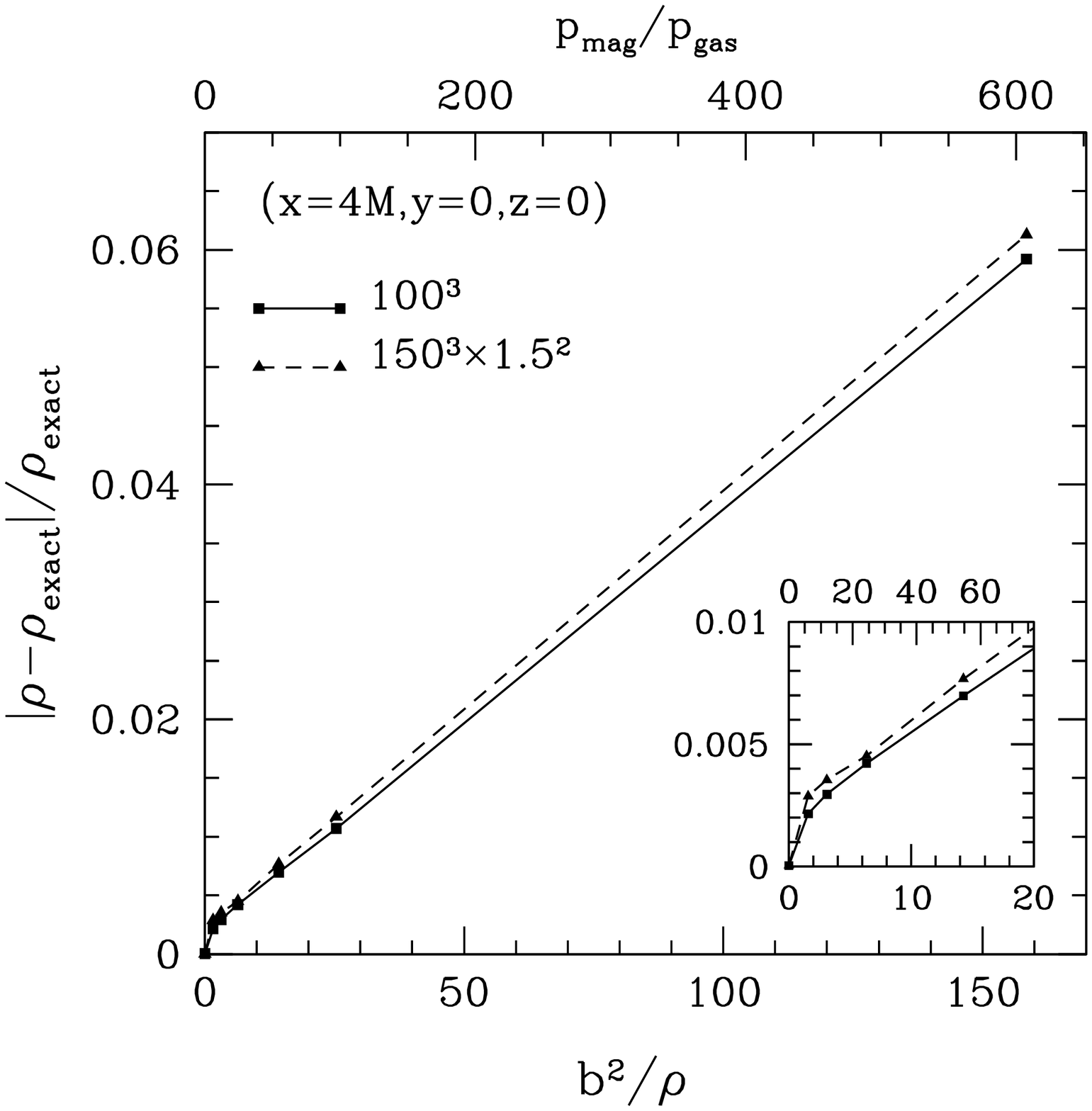}
  \end{center}
  \caption{\label{fig_whiskymhd:michel}\textit{Left panel:} $L_1$ norm
    of the relative error in the rest-mass density for the magnetized
    spherical accretion test, shown for different values of the
    magnetic field. Results from $100^3$ and $150^3$ runs are compared
    at time $t=100M$, with the high-resolution curve being multiplied
    by $1.5^2$ so that with a second-order convergence the two lines
    would overlap. \textit{Right panel:} Relative error computed at
    point $x=4M,\,y=0,\,z=0$; also in this case the high-resolution
    line is scaled to produce an overlap in the case of second-order
    convergence. In both cases the insets offer a magnification for
    small values of the magnetic field.}
\end{figure}

We measure the order of accuracy of the code by using the $L_1$-norm
of the relative error on the rest-mass density
\begin{equation}
\frac{||\delta\rho||_1}{||\rho||_1} \equiv 
	\frac{\sum_{i,j,k}\left|\rho_{i,j,k}-
        \rho_{\rm exact}(x_i,y_j,z_k)\right|}{\sum_{i,j,k}
	\rho_{\rm exact}(x_i,y_j,z_k)} \; .
\end{equation}
We plot this quantity in the left panel of~\fref{fig_whiskymhd:michel}
as a function of the magnetic-field strength and as computed at time
$t=100M$ for two different resolutions of $100^3$ and $150^3$
gridpoints, respectively. In addition, the error from the
high-resolution simulation is multiplied by $1.5^2$ so that the two
curves should overlap if the code were second-order
convergent. Clearly, the code does not show the expected convergence
rate but for relatively weak magnetizations, i.e. $b^2/\rho \lesssim
4$ (we recall that these corresponds nevertheless to rather large
magnetic fields of $\approx 10^{19}$ G). This behaviour is indeed
similar to what found by Duez {\it et al.}~\cite{duez05} and has a
rather simple explanation. It is due to the rather large error in
gridpoints near the excision boundary, \textit{i.e.,} for $x^i \in
[M,2M]$, which spoil the overall behaviour of the $L_1$ norm
(admittedly not a good measure of the convergence for a solution which
is so rapidly varying near the excision boundary). To clarify this, we
show in the right panel of~\fref{fig_whiskymhd:michel} the same as in
the left panel but for the relative error computed at a single
gridpoint, {\it i.e.,} at $x=4M,y=0,z=0$. Clearly the convergence is
much closer to second-order in this case (the precise order being
$\approx 1.8$) and for much larger range in magnetizations.

\begin{figure}[h]
  \begin{center}
  \includegraphics[width=0.6\textwidth]{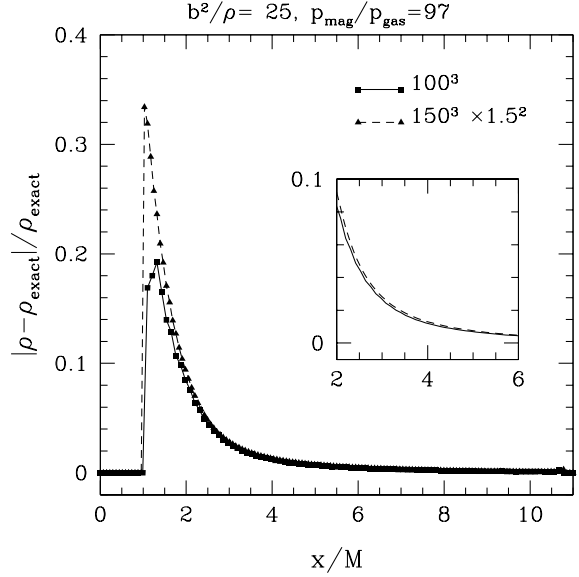}
  \end{center}
  \caption{\label{fig_whiskymhd:michel_x} Relative error of the
    rest-mass density for the magnetized spherical-accretion flow with
    $b^2/\rho=25$ (i.e., with $p_{mag}/p_{gas}=97$) along the $x$-axis
    at $t=100M$. The high-resolution error is multiplied by $1.5^2$ so
    that the two lines would overlap with a second-order
    convergence. Clearly, this does not happen near the excision
    boundary located at $x=M$.}
\end{figure}

A closer look at the behaviour of the relative error is offered
in~\fref{fig_whiskymhd:michel_x}, where it is shown as measured along
the $x$-direction for a magnetization of $b^2/\rho=25$ (i.e., with
$p_{mag}/p_{gas}=97$) and at time $t=100M$. Also in this case, the
high-resolution relative error is multiplied by $1.5^2$ so that the
two lines overlap if second-order convergent. Clearly this does not
happen but also only for a small number of gridpoints near the
excision boundary located at $x=M$.

As a final remark, we point out that the simulations of
spherical-accretion flows performed here span a range of
magnetizations well beyond what considered in the past with codes
using Cartesian coordinates (the results reported in~\cite{duez05},
for instance, were limited to $b^2/\rho \lesssim 30$). Indeed, no sign
of instability has been found and only a moderate loss of accuracy has
been measured for magnetic fields as large as $b^2/\rho \lesssim 160$.

\subsection{\label{sec:magstar}Evolution of a stable magnetized neutron star}

As a final test validating the code in a fully dynamical evolution of
both the MHD variables and of the spacetime, we now consider the
evolution of a stable magnetized neutron star. Although this initial
data represents a stationary solution, small oscillations can be
triggered by the small but nonzero truncation error. Such oscillations
are sometimes considered a nuisance and even suppressed through the
introduction of artificial-viscosity terms. On the contrary, since
they represent the consistent response of the star to small
perturbations, they should considered as extremely useful.  The
eigenfunctions and eigenfrequencies of these oscillations, in fact,
can serve both as a test of the code, when compared with the
expectations coming from perturbation theory (see Appendix B
of~\cite{whisky} for a representative example), or to extract
information on the properties of the star, when considering regimes
which are not yet accessible to perturbative studies (e.g., in the
case of nonlinear oscillations or very rapidly rotating stars).

Two options are possible for the construction of the initial data. A
first and simpler one was employed extensively
in~\cite{duez06a,duez06b,duez06c} and consists of considering a
background purely-hydrodynamical solution in stable dynamical
equilibrium and of ``adding'' a poloidal magnetic field in terms of a
purely toroidal vector potential. Besides being essentially arbitrary,
the vector potential is chosen to be proportional to the pressure so
as to lead to a magnetic field entirely confined within the
star. While straightforward, this approach does not construct a
magnetized stellar model which is consistent solution of the Einstein
equations and thus inevitably introduces violations of the Hamiltonian
and momentum constraints. Such violations, however, are in general
negligible for reasonably small magnetic fields.

\begin{figure}[h]
  \begin{center}
    \includegraphics[angle=-90,width=0.6\textwidth]{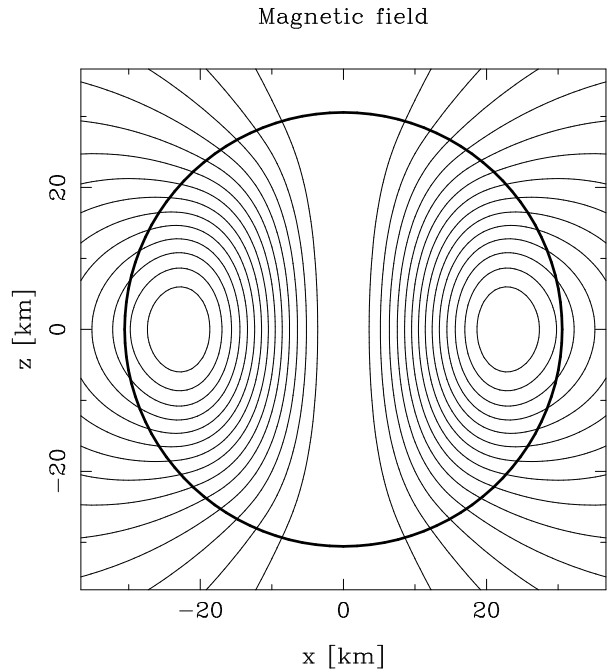}
  \end{center}
  \caption{\label{fig_whiskymhd:magstar_C1_1e-6}Magnetic field lines
    of the oscillating magnetized and nonrotating neutron star
    considered in this paper. The solid thick line represents the star
    surface.}
\end{figure}

A second option, and the one employed here, consists of computing the
initial conditions as a consistent and accurate solution of the
Einstein equations for a stationary, axisymmetric and magnetized
star. We have done this by using the spectral code developed by
Bocquet {\it et al.}~\cite{novak}, which solves the full set of
Einstein and Maxwell equations to high precision.  Assuming an
axisymmetric model with a poloidal magnetic field having the dipole
moment aligned along the rotation axis, the code is used to build
initial configurations of uniformly rotating magnetized neutron stars
with different angular velocities and magnetic field strengths.

For simplicity we here consider a nonrotating magnetized neutron star
with mass $M=1.3M_\odot$ endowed with a poloidal magnetic field with
magnetic dipole along the $z$-axis and a central magnetic field
$B_c=2.4\times 10^{14}\,$G, corresponding to $\beta=p_{\rm mag}/p_{\rm
  gas}=10^{-6}$ (this $\beta$, which should not be confused with the
shift vector $\beta^i$, is always monotonically decreasing inside the
initial equilibrium model, and is much larger in the atmosphere, where
it reaches values $\sim 10^6$). A polytropic equation of state with
$\Gamma=2$ and $K=372$ was used both for the calculation of the
initial model and during the evolution. A representative image of the
initial model is presented in~\fref{fig_whiskymhd:magstar_C1_1e-6},
which shows the magnetic field lines together with the stellar surface
(thick solid line). Note that although the star is nonrotating, the
presence of a magnetic field replaces the spherical symmetry for an
axisymmetrical one.

\begin{figure}
  \begin{center}
    \includegraphics[width=0.49\textwidth]{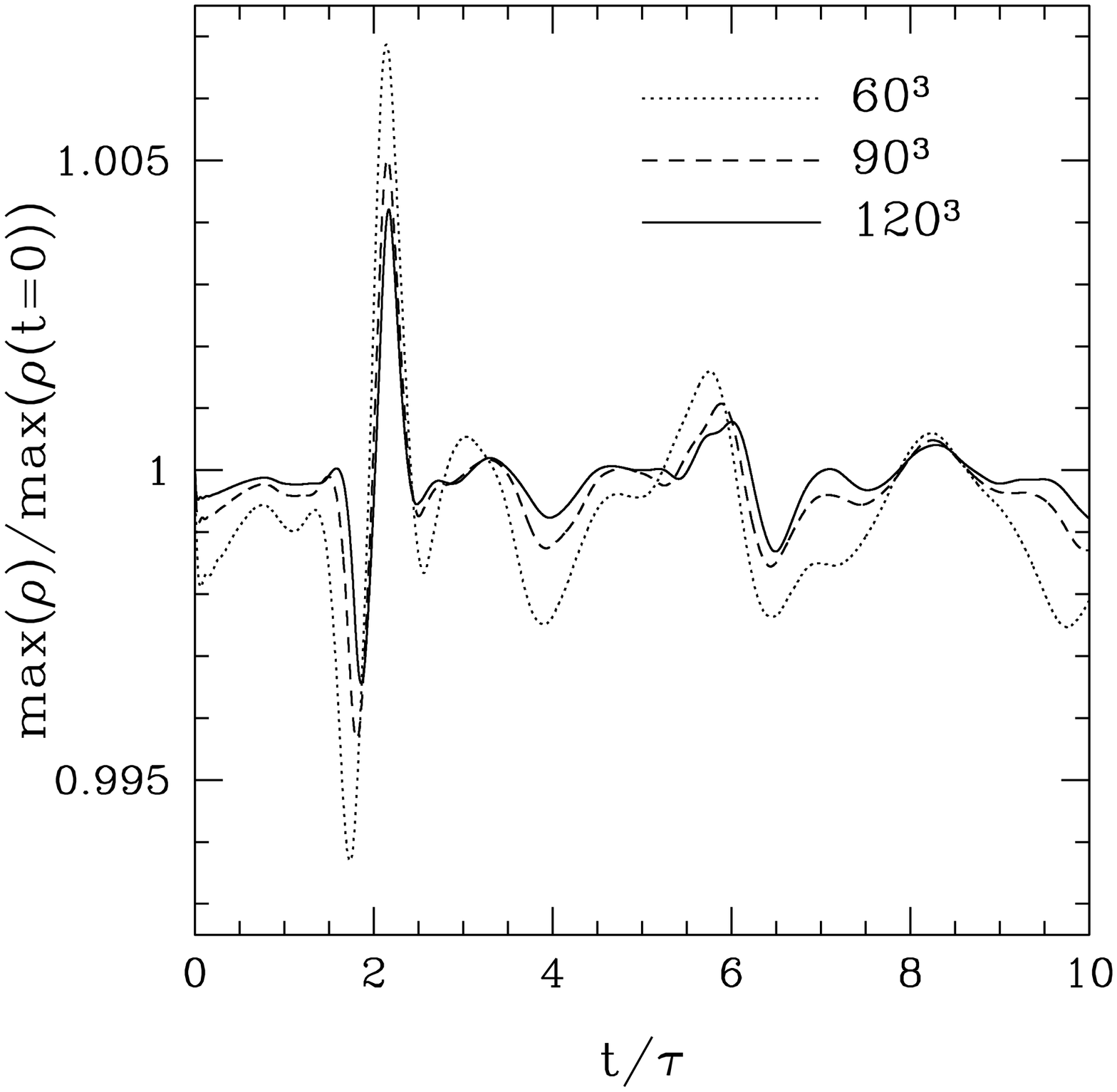}
    \hfill
    \includegraphics[width=0.49\textwidth]{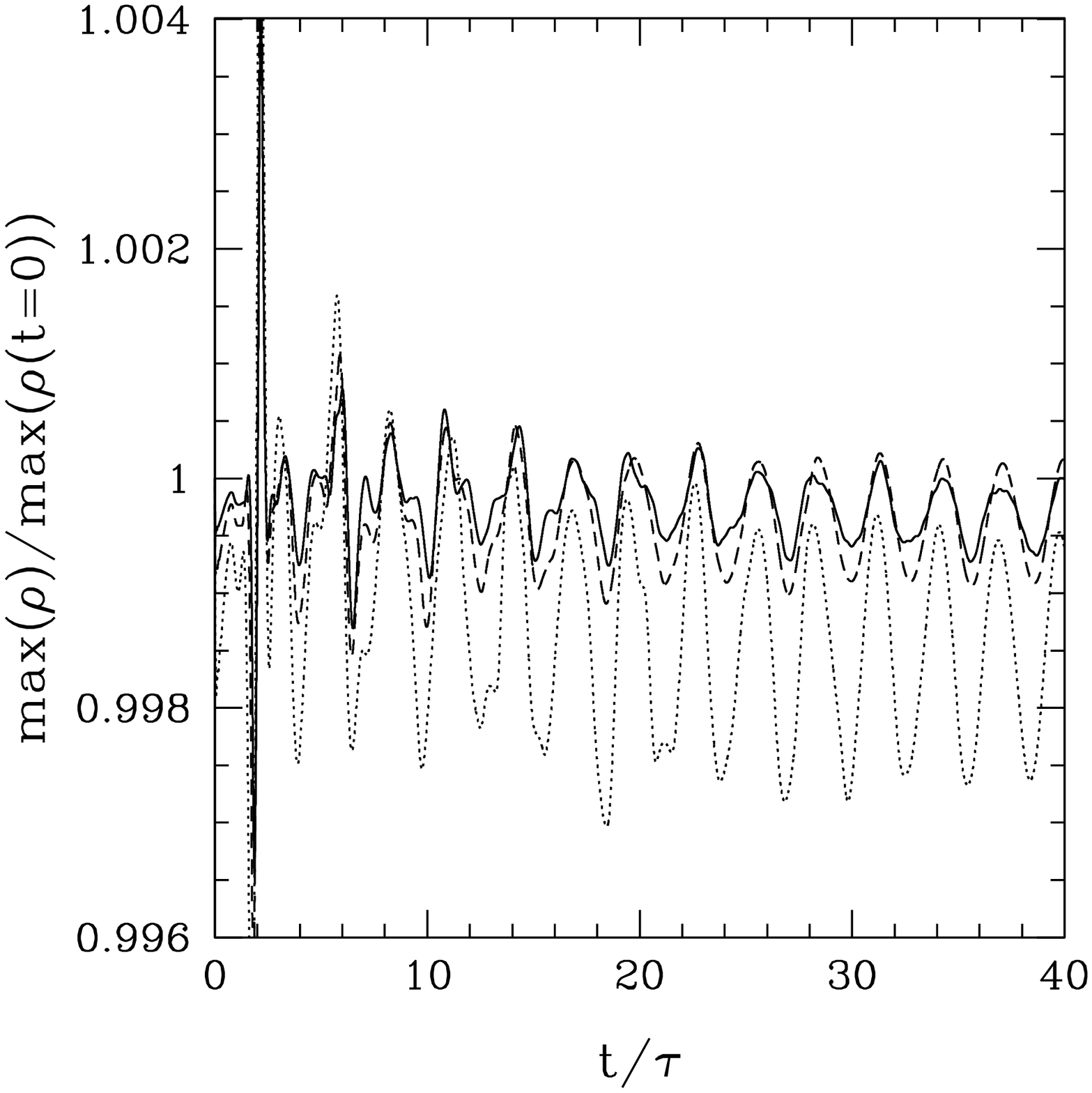}
    \caption{\label{fig_whiskymhd:rhomax}\textit{Left panel:} Maximum
      of the rest-mass density $\rho$ normalized to its initial value
      and expressed in terms of the dynamical timescale
      $\tau\equiv\sqrt{R^3/M}$. The magnetized star is evolved within
      the Cowling approximation, with different lines referring to
      different resolutions: $N=60^3$ (dotted), $N=90^3$ (dashed line)
      and $N=120^3$ (solid line). \textit{Right panel:} The same as in
      the left one but for a longer timescale.}
  \end{center}
\end{figure}

As a first test, we consider the evolution of the star within the
so-called ``Cowling approximation'', {\it i.e.} by holding the metric
fixed to its initial value and by evolving the MHD variables onto this
background spacetime (the evolution is not made only at the outer
boundaries, where we use Dirichlet-type boundary conditions). The
results of these evolutions are summarized
in~\fref{fig_whiskymhd:rhomax}, with the left panel showing the
evolution of the maximum of the rest-mass density $\rho$ when
normalized to its initial value. The three different lines (dotted,
dashed and continuous) refer to the three resolutions used of
$N=60^3,\, 90^3$, and $120^3$, respectively. The coordinate time on
the horizontal axis is expressed in terms of the characteristic
``dynamical timescale'' $\tau\equiv\sqrt{R^3/M}$, where $R$ is the
coordinate radius of the star.

In analogy with what observed in the purely hydrodynamical
case~\cite{whisky}, the magnetized star is set into oscillation by the
small truncation error introduced by the mapping onto a Cartesian
coordinate system of the stationary solution found in spherical polar
coordinates~\cite{novak} (no perturbation coming from the outer
boundaries was seen to influence the dynamics of the
oscillations). Because of its stochastic nature, the initial
perturbation triggers a variety of modes, most of which however decay
rather rapidly leaving, after $t \simeq 25 \tau$, an essentially
harmonic oscillation in the fundamental mode only. This is shown in
the right panel of~\fref{fig_whiskymhd:rhomax} which shows the
evolution over a longer timescale.  Furthermore, in the linear regime
considered here, the amplitude of the oscillations is proportional to
the magnitude of the truncation error and one expects the former to
decrease as the resolution is increased. This is clearly the case for
the oscillations shown in~\fref{fig_whiskymhd:rhomax}, and the very
good overlap among the different timeseries is an indication that the
oscillations indeed correspond to eigenmodes and that the code is
tracking them correctly at these resolutions.

\begin{figure}
  \begin{center}
    \includegraphics[width=0.49\textwidth]{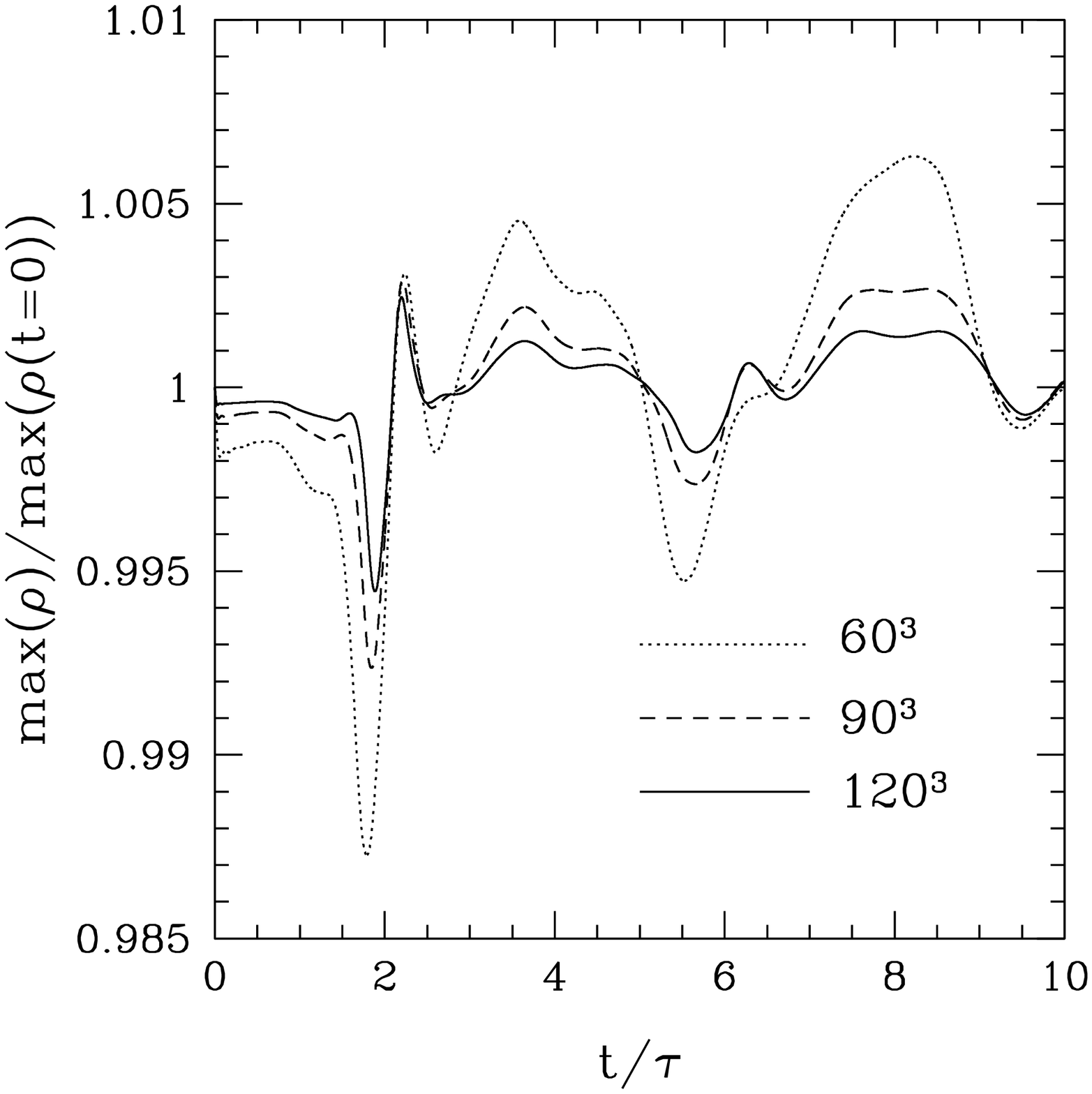}
    \hfill
    \includegraphics[width=0.49\textwidth]{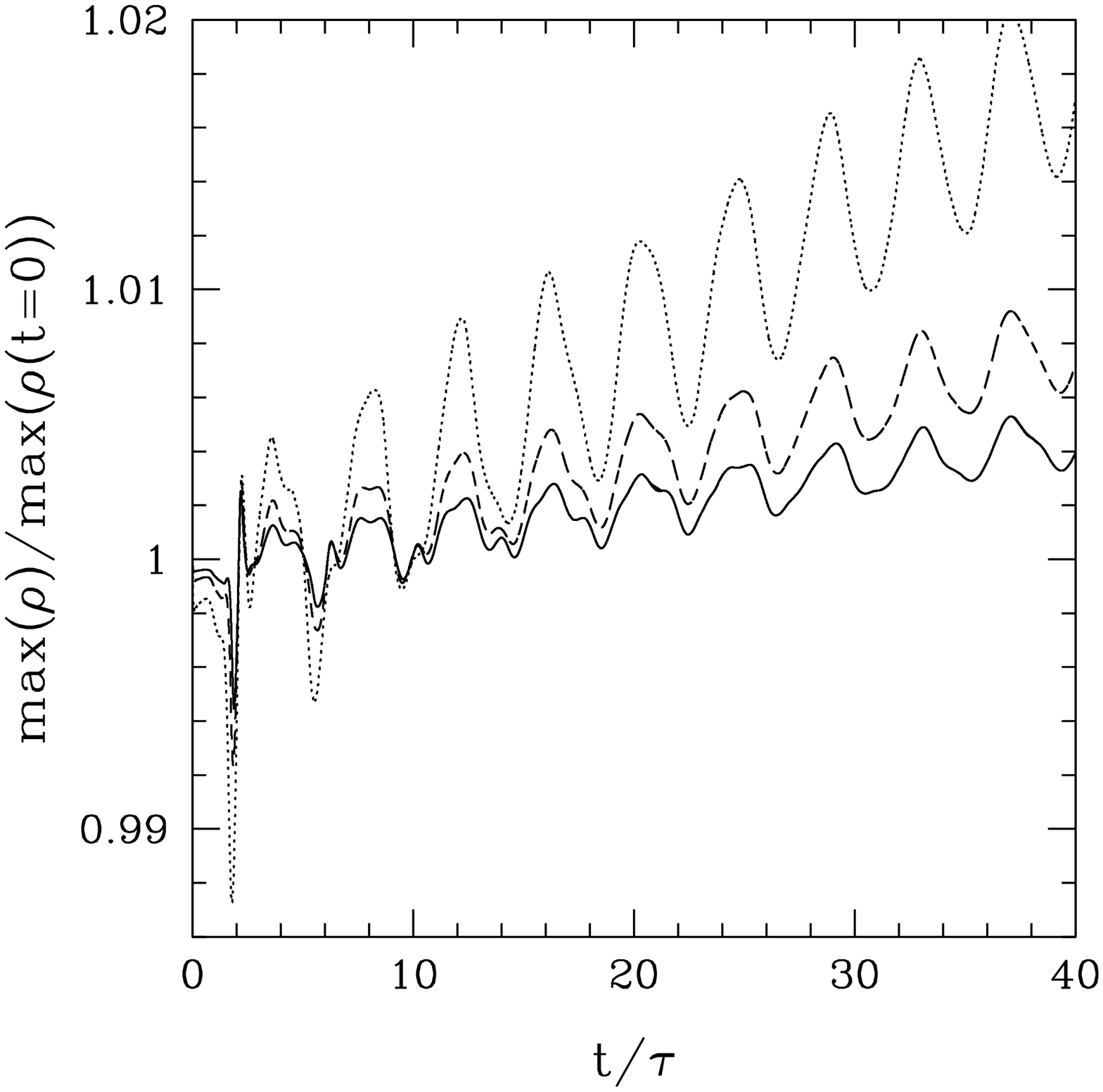}
  \end{center}
    \caption{\label{fig_whiskymhd:rhomax_fullgr}\textit{Left panel:}
      Maximum of the rest-mass density $\rho$ normalized to its
      initial value and expressed in terms of the dynamical timescale
      $\tau\equiv\sqrt{R^3/M}$. The magnetized star is evolved
      together with the spacetime, with different lines referring to
      different resolutions: $N=60^3$ (dotted), $N=90^3$ (dashed line)
      and $N=120^3$ (solid line). \textit{Right panel:} The same as in
      the left one but for a longer timescale.}
\end{figure}

Next, we consider the evolution of the same initial model discussed
above but including also the solution of the Einstein equations so as
to make the system fully dynamical (Dirichlet-type boundary conditions
are used at the outer boundaries for the MHD variables and radiative
ones for the fields). Also in this case, oscillations are triggered by
the truncation error, with an amplitude that converges to zero with
the increase of the resolution and with an harmonic content that
becomes more evident after the initial transient (also in this case,
no perturbation coming from the outer boundaries was seen to influence
the dynamics of the oscillations). In addition, and in analogy with
what observed in the purely hydrodynamical
case~\cite{Font99,Font02c,whisky}, the oscillations are accompanied by
a secular growth which also converges away at the correct rate with
increasing grid resolution and that does not influence the long-term
evolutions. This is shown in~\fref{fig_whiskymhd:rhomax_fullgr} which
reports the same quantities as in~\fref{fig_whiskymhd:rhomax} but for
a fully dynamical evolution. Note also that the secular evolution of
the central rest-mass density varies according to the EOS adopted:
when using the ideal-fluid EOS, in fact, the secular drift of the
central rest-mass density is towards lower densities. However, if the
adiabatic condition is enforced, the opposite is true and central
rest-mass density evolves towards larger values. Both the evolutions
in the Cowling approximation and in dynamical spacetimes, have not
shown signs of instability at all resolutions considered and up to
several tens of dynamical timescales.

As a final remark we underline that the convergence rate is not
exactly second-order but slightly smaller, (i.e., 1.7-1.8), because
the reconstruction schemes are only first-order accurate at local
extrema ({\it i.e.}  at the centre and at the surface of the star)
thus increasing the overall truncation error. Similar estimates were
obtained also in the purely hydrodynamical case~\cite{whisky}.

\section{\label{conclusion}Conclusions}

We have presented a new three-dimensional numerical code in Cartesian
coordinates developed to solve the full set of GRMHD equation on a
dynamical background. The code is based on high-resolution
shock-capturing techniques as implemented on domains with adaptive
mesh refinements. This code represents the extension to MHD of the
approach already implemented with success in the general-relativistic
hydrodynamics code \texttt{Whisky}~\cite{Baiotti03a}.

The code has been validated through an extensive series of testbeds
both in special and in general relativity scenarios. In particular, we
have first considered a set of Riemann problems in a Minkowski
spacetime following a variety of initial conditions. In all of the
tests presented, the numerical solution has been compared with the
exact one~\cite{giacomazzo06}, providing, for the first time, a
quantitative assessment of the code's ability to evolve correctly and
accurately all of the different waves that can form in relativistic
MHD. Furthermore, as a demonstration of the proper handling of
continuous and discontinuous flows in the presence of an excision
region, we have extended the Riemann-problems tests across an excised
boundary. In doing so we have revealed the importance of correct
boundary conditions and pointed out that those recently proposed
in~\cite{duez05} can lead to incorrect solutions for non-smooth flows.

Next, to investigate magnetized fluids in a curved but fixed
spacetime, we have considered the spherical accretion of a perfect
fluid with a radial magnetic field onto a Schwarzschild black hole
(relativistic Bondi flow). Also in this case, the code has been shown
to accurately reproduce the stationary solution and to be convergent
at the correct rate for small and large magnetizations. For very large
magnetizations, however, the very rapidly varying behaviour of the MHD
variables near the excision region prevents from a correct convergence
near the horizon, although the code remains second-convergent away
from the horizon and is convergent overall. Also for these
extremely-high values of the magnetic field, the code has shown to be
robust and accurate at regimes where other codes were reported to
fail~\cite{duez05}.

Finally, we have considered the evolution of magnetized neutron stars
in equilibrium and constructed as a consistent solutions of the
coupled Einstein-Maxwell equations. Such initial models represent an
important difference from those considered by other authors, which
were not consistent solution of the Einstein equations initially and
whose magnetic field is totally confined within the
star~\cite{duez05}. In analogy with what observed in the purely
hydrodynamical case~\cite{whisky}, these magnetized stars are set into
oscillation by the small truncation error. These pulsations, which
have been studied both in fixed (Cowling approximation) and in
dynamical spacetimes, have been shown to have the correct behaviour
under changes of spatial resolution and to correspond to the
eigenmodes of relativistic and magnetized stars. Both evolutions in
the Cowling approximation and in dynamical spacetimes have not shown
signs of instability at all resolutions considered and up to several
tens of dynamical timescales.

A number of projects are expected to be carried out with the new
code. Firstly, we plan to extend the study on the oscillations of
rotating and nonrotating neutron stars with a detailed analysis of the
effect of magnetic fields on the frequency of oscillations. It is
important to note that only recently some results were obtained in
perturbation theory and within the Cowling
approximation~\cite{sotani}. Our code will be a complementary tool to
the perturbative approaches, using the latter as testbeds and carrying
them beyond the regimes of slow rotation and weak magnetizations. Such
a study, and the comparison with the frequencies observed in objects
such as the soft gamma-repeaters, will provide useful information on
the mass and magnetic-field strength of magnetars.

Secondly, we plan to use \texttt{WhiskyMHD} to study the collapse of
uniformly and differentially rotating magnetized neutron stars with
the aim of extending further the work done
in~\cite{whisky,baiotti05,baiotti06} and to highlight the role that
this process may have in the phenomenology of short $\gamma$-ray
bursts. We are especially interested in the calculation of the
gravitational-wave signal emitted by these sources and on the
influence that magnetic fields may have on it.

Finally, and in view of the total generality with which it has been
developed, the code will be used to study the dynamics of binary
systems of neutron stars and mixed binaries, with the aim of extending
the work carried out in~\cite{loeffler06} and of considering in a
fully general-relativistic context the Newtonian results obtained
in~\cite{price06}.

\ack 

It is a pleasure to thank L. Baiotti, G. Bodo, L. Del Zanna, I. Hawke,
F. L\"offler, C.~D. Ott, E. Schnetter and O. Zanotti for useful
discussions and comments. We are also indebted with J. Novak for his
help in the construction of the equilibrium magnetized stars. All the
numerical computations were performed on clusters {\it Albert2} at the
Physics Department of the University of Parma (Parma, Italy), {\it
CLX} at CINECA (Bologna, Italy) and {\it Peyote} at the AEI (Golm,
Germany).


\section*{References}
\bibliographystyle{unsrt}

\bibliography{whiskymhd_cqg}

\end{document}